\documentclass[%
 twocolumn,longbibliography,
 aps,prx,
 amsmath,amssymb,
 superscriptaddress,
 english,
 nofootinbib,
 reprint,%
]{revtex4-1}
\usepackage{stmaryrd}
\usepackage{color}
\usepackage{array}
\usepackage{enumitem}
\usepackage{mathpazo}
\usepackage{setspace}

\usepackage{bm}
\usepackage{amssymb}
\usepackage{amsthm}
\usepackage{mathtools}
\usepackage{multirow}
\usepackage{cases}
\usepackage[colorlinks=true,linkcolor=blue,citecolor=blue,pdfauthor={ },pdftitle={ },pdfsubject={ },pdfkeywords={ }]{hyperref}
\usepackage{cleveref}
\usepackage{pgfplots}
\usepackage{lipsum}
\usepackage{youngtab}
\usepackage{subfigure}
\usepackage{graphicx}

\newtheorem{definition}{Definition}
\newtheorem{proposition}[definition]{Proposition}
\newtheorem{lemma}[definition]{Lemma}

\newtheorem{theorem}[definition]{Theorem}

\newtheorem{corollary}[definition]{Corollary}
\newtheorem{conjecture}[definition]{Conjecture}

\newtheorem{remark}[definition]{Remark}
\newtheorem{example}[definition]{Example}
\newtheorem{question}[definition]{Question}

\def\Dbar{\leavevmode\lower.6ex\hbox to 0pt
{\hskip-.23ex\accent"16\hss}D}
\makeatletter
\def\url@leostyle{%
  \@ifundefined{selectfont}{\def\UrlFont{\sf}}{\def\UrlFont{\small\ttfamily}}}
\makeatother
\DeclareMathOperator{\Tr}{Tr} %

\def\bcj{\begin{conjecture}}
\def\ecj{\end{conjecture}}
\def\bcr{\begin{corollary}}
\def\ecr{\end{corollary}}
\def\bd{\begin{definition}}
\def\ed{\end{definition}}
\def\bea{\begin{eqnarray}}
\def\eea{\end{eqnarray}}
\def\bem{\begin{enumerate}}
\def\eem{\end{enumerate}}
\def\bex{\begin{example}}
\def\eex{\end{example}}
\def\bim{\begin{itemize}}
\def\eim{\end{itemize}}
\def\bl{\begin{lemma}}
\def\el{\end{lemma}}
\def\bpf{\begin{proof}}
\def\epf{\end{proof}}
\def\bpp{\begin{proposition}}
\def\epp{\end{proposition}}
\def\bqu{\begin{question}}
\def\equ{\end{question}}
\def\br{\begin{remark}}
\def\er{\end{remark}}
\def\bt{\begin{theorem}}
\def\et{\end{theorem}}

\def\btb{\begin{tabular}}
\def\etb{\end{tabular}}

\newcommand{\nc}{\newcommand}

 \nc{\bA}{{\bf A}} \nc{\bB}{{\bf B}} \nc{\bC}{{\bf C}}
 \nc{\bD}{{\bf D}} \nc{\bE}{{\bf E}} \nc{\bF}{{\bf F}}
 \nc{\bG}{{\bf G}} \nc{\bH}{{\bf H}} \nc{\bI}{{\bf I}}
 \nc{\bJ}{{\bf J}} \nc{\bK}{{\bf K}} \nc{\bL}{{\bf L}}
 \nc{\bM}{{\bf M}} \nc{\bN}{{\bf N}} \nc{\bO}{{\bf O}}
 \nc{\bP}{{\bf P}} \nc{\bQ}{{\bf Q}} \nc{\bR}{{\bf R}}
 \nc{\bS}{{\bf S}} \nc{\bT}{{\bf T}} \nc{\bU}{{\bf U}}
 \nc{\bV}{{\bf V}} \nc{\bW}{{\bf W}} \nc{\bX}{{\bf X}}
 \nc{\bZ}{{\bf Z}}

\nc{\cA}{{\cal A}} \nc{\cB}{{\cal B}} \nc{\cC}{{\cal C}}
\nc{\cD}{{\cal D}} \nc{\cE}{{\cal E}} \nc{\cF}{{\cal F}}
\nc{\cG}{{\cal G}} \nc{\cH}{{\cal H}} \nc{\cI}{{\cal I}}
\nc{\cJ}{{\cal J}} \nc{\cK}{{\cal K}} \nc{\cL}{{\cal L}}
\nc{\cM}{{\cal M}} \nc{\cN}{{\cal N}} \nc{\cO}{{\cal O}}
\nc{\cP}{{\cal P}} \nc{\cQ}{{\cal Q}} \nc{\cR}{{\cal R}}
\nc{\cS}{{\cal S}} \nc{\cT}{{\cal T}} \nc{\cU}{{\cal U}}
\nc{\cV}{{\cal V}} \nc{\cW}{{\cal W}} \nc{\cX}{{\cal X}}
\nc{\cZ}{{\cal Z}}

\nc{\hA}{{\hat{A}}} \nc{\hB}{{\hat{B}}} \nc{\hC}{{\hat{C}}}
\nc{\hD}{{\hat{D}}} \nc{\hE}{{\hat{E}}} \nc{\hF}{{\hat{F}}}
\nc{\hG}{{\hat{G}}} \nc{\hH}{{\hat{H}}} \nc{\hI}{{\hat{I}}}
\nc{\hJ}{{\hat{J}}} \nc{\hK}{{\hat{K}}} \nc{\hL}{{\hat{L}}}
\nc{\hM}{{\hat{M}}} \nc{\hN}{{\hat{N}}} \nc{\hO}{{\hat{O}}}
\nc{\hP}{{\hat{P}}} \nc{\hR}{{\hat{R}}} \nc{\hS}{{\hat{S}}}
\nc{\hT}{{\hat{T}}} \nc{\hU}{{\hat{U}}} \nc{\hV}{{\hat{V}}}
\nc{\hW}{{\hat{W}}} \nc{\hX}{{\hat{X}}} \nc{\hZ}{{\hat{Z}}}

\newcommand{\bra}[1]{\langle#1|}
\newcommand{\ket}[1]{|#1\rangle}

\newcommand{\ketbra}[2]{|#1\rangle\!\langle#2|}

\def\Dbar{\leavevmode\lower.6ex\hbox to 0pt
{\hskip-.23ex\accent"16\hss}D}
\begin{document}


\def\be{\begin{eqnarray}}
\def\ee{\end{eqnarray}}


\newcommand{\ca}{\mathcal A}

\newcommand{\cb}{\mathcal B}
\newcommand{\cc}{\mathcal C}
\newcommand{\cd}{\mathcal D}
\newcommand{\ce}{\mathcal E}
\newcommand{\cf}{\mathcal F}
\newcommand{\cg}{\mathcal G}
\newcommand{\ch}{\mathcal H}
\newcommand{\ci}{\mathcal I}
\newcommand{\cj}{\mathcal J}
\newcommand{\ck}{\mathcal K}
\newcommand{\cl}{\mathcal L}
\newcommand{\cm}{\mathcal M}
\newcommand{\cn}{\mathcal N}
\newcommand{\co}{\mathcal O}
\newcommand{\cp}{\mathcal P}
\newcommand{\cq}{\mathcal Q}
\newcommand{\calr}{\mathcal R}
\newcommand{\cs}{\mathcal S}
\newcommand{\ct}{\mathcal T}
\newcommand{\cu}{\mathcal U}
\newcommand{\cv}{\mathcal V}
\newcommand{\cw}{\mathcal W}
\newcommand{\cx}{\mathcal X}
\newcommand{\cy}{\mathcal Y}
\newcommand{\cz}{\mathcal Z}


\newcommand{\sa}{\mathscr{A}}
\newcommand{\sm}{\mathscr{M}}


\newcommand{\fa}{\mathfrak{a}}  \newcommand{\Fa}{\mathfrak{A}}
\newcommand{\fb}{\mathfrak{b}}  \newcommand{\Fb}{\mathfrak{B}}
\newcommand{\fc}{\mathfrak{c}}  \newcommand{\Fc}{\mathfrak{C}}
\newcommand{\fd}{\mathfrak{d}}  \newcommand{\Fd}{\mathfrak{D}}
\newcommand{\fe}{\mathfrak{e}}  \newcommand{\Fe}{\mathfrak{E}}
\newcommand{\ff}{\mathfrak{f}}  \newcommand{\Ff}{\mathfrak{F}}
\newcommand{\fg}{\mathfrak{g}}  \newcommand{\Fg}{\mathfrak{G}}
\newcommand{\fh}{\mathfrak{h}}  \newcommand{\Fh}{\mathfrak{H}}
\newcommand{\fraki}{\mathfrak{i}}       \newcommand{\Fraki}{\mathfrak{I}}
\newcommand{\fj}{\mathfrak{j}}  \newcommand{\Fj}{\mathfrak{J}}
\newcommand{\fk}{\mathfrak{k}}  \newcommand{\Fk}{\mathfrak{K}}
\newcommand{\fl}{\mathfrak{l}}  \newcommand{\Fl}{\mathfrak{L}}
\newcommand{\fm}{\mathfrak{m}}  \newcommand{\Fm}{\mathfrak{M}}
\newcommand{\fn}{\mathfrak{n}}  \newcommand{\Fn}{\mathfrak{N}}
\newcommand{\fo}{\mathfrak{o}}  \newcommand{\Fo}{\mathfrak{O}}
\newcommand{\fp}{\mathfrak{p}}  \newcommand{\Fp}{\mathfrak{P}}
\newcommand{\fq}{\mathfrak{q}}  \newcommand{\Fq}{\mathfrak{Q}}
\newcommand{\fr}{\mathfrak{r}}  \newcommand{\Fr}{\mathfrak{R}}
\newcommand{\fs}{\mathfrak{s}}  \newcommand{\Fs}{\mathfrak{S}}
\newcommand{\ft}{\mathfrak{t}}  \newcommand{\Ft}{\mathfrak{T}}
\newcommand{\fu}{\mathfrak{u}}  \newcommand{\Fu}{\mathfrak{U}}
\newcommand{\fv}{\mathfrak{v}}  \newcommand{\Fv}{\mathfrak{V}}
\newcommand{\fw}{\mathfrak{w}}  \newcommand{\Fw}{\mathfrak{W}}
\newcommand{\fx}{\mathfrak{x}}  \newcommand{\Fx}{\mathfrak{X}}
\newcommand{\fy}{\mathfrak{y}}  \newcommand{\Fy}{\mathfrak{Y}}
\newcommand{\fz}{\mathfrak{z}}  \newcommand{\Fz}{\mathfrak{Z}}

\newcommand{\cfg}{\dot \fg}
\newcommand{\cFg}{\dot \Fg}
\newcommand{\ccg}{\dot \cg}
\newcommand{\circj}{\dot {\mathbf J}}
\newcommand{\circs}{\circledS}
\newcommand{\jmot}{\mathbf J^{-1}}


\newcommand{\rmd}{\mathrm d}
\newcommand{\mca}{\ ^-\!\!\ca}
\newcommand{\pca}{\ ^+\!\!\ca}
\newcommand{\peq}{^\Psi\!\!\!\!\!=}
\newcommand{\lt}{\left}
\newcommand{\rt}{\right}
\newcommand{\HN}{\hat{H}(N)}
\newcommand{\HM}{\hat{H}(M)}
\newcommand{\Hv}{\hat{H}_v}
\newcommand{\cyl}{\mathbf{Cyl}}
\newcommand{\lag}{\left\langle}
\newcommand{\rag}{\right\rangle}
\newcommand{\Ad}{\mathrm{Ad}}
\newcommand{\trace}{\mathrm{tr}}
\newcommand{\bbc}{\mathbb{C}}
\newcommand{\AC}{\overline{\mathcal{A}}^{\mathbb{C}}}
\newcommand{\Ar}{\mathbf{Ar}}
\newcommand{\uc}{\mathrm{U(1)}^3}
\newcommand{\M}{\hat{\mathbf{M}}}
\newcommand{\spin}{\text{Spin(4)}}
\newcommand{\id}{\mathrm{id}}
\newcommand{\Pol}{\mathrm{Pol}}
\newcommand{\Fun}{\mathrm{Fun}}
\newcommand{\bp}{p}
\newcommand{\act}{\rhd}
\newcommand{\data}{\lt(j_{ab},A,\bar{A},\xi_{ab},z_{ab}\rt)}
\newcommand{\datao}{\lt(j^{(0)}_{ab},A^{(0)},\bar{A}^{(0)},\xi_{ab}^{(0)},z_{ab}^{(0)}\rt)}
\newcommand{\deltadata}{\lt(j'_{ab}, A',\bar{A}',\xi_{ab}',z_{ab}'\rt)}
\newcommand{\background}{\lt(j_{ab}^{(0)},g_a^{(0)},\xi_{ab}^{(0)},z_{ab}^{(0)}\rt)}
\newcommand{\sgn}{\mathrm{sgn}}
\newcommand{\vth}{\vartheta}
\newcommand{\rmi}{\mathrm{i}}
\newcommand{\bfmu}{\pmb{\mu}}
\newcommand{\bfnu}{\pmb{\nu}}
\newcommand{\bfm}{\mathbf{m}}
\newcommand{\bfn}{\mathbf{n}}
\newcommand{\perk}{\mathfrak{S}_k}
\newcommand{\dens}{\mathrm{D}}
\newcommand{\iden}{\mathbb{I}}
\newcommand{\End}{\mathrm{End}}


\newcommand{\sz}{\mathscr{Z}}
\newcommand{\sk}{\mathscr{K}}

\title{Simulating Noisy Quantum Circuits with Matrix Product Density Operators}

\author{Song Cheng}
\affiliation{Yanqi Lake Beijing Institute of Mathematical Sciences and Applications, Beijing, 100407, China}
\affiliation{Peng Cheng Laboratory, Shenzhen, 518055, China}
\author{Chenfeng Cao}
\affiliation{Department of Physics, The Hong Kong University of Science and Technology, Clear Water Bay, Kowloon, Hong Kong, China}
\author{Chao Zhang}
\affiliation{Peng Cheng Laboratory, Shenzhen, 518055, China}
\author{Yongxiang Liu}
\affiliation{Peng Cheng Laboratory, Shenzhen, 518055, China}
\author{Shi-Yao Hou}
\affiliation{College of Physics and Electronic Engineering, Center for Computational Sciences,  Sichuan Normal University, Chengdu 610068, China}
\affiliation{Peng Cheng Laboratory, Shenzhen, 518055, China}
\author{Pengxiang Xu}
\affiliation{Peng Cheng Laboratory, Shenzhen, 518055, China}
\author{Bei Zeng}
\email{zengb@ust.hk}
\affiliation{Department of Physics, The Hong Kong University of Science and Technology, Clear Water Bay, Kowloon, Hong Kong, China}
\date{\today}

\begin{abstract}
Simulating quantum circuits with classical computers requires resources growing exponentially
in terms of system size. Real quantum computer with noise, however, may be simulated polynomially with various methods considering different noise models. In this work, we simulate random quantum circuits in 1D with Matrix Product Density Operators (MPDO), for different noise models such as dephasing, depolarizing, and amplitude damping. We show that the method based on Matrix Product States (MPS) fails to approximate the noisy output quantum states for any of the noise models considered, while the MPDO method approximates them well. Compared with the method of Matrix Product Operators (MPO), the MPDO method reflects a clear physical picture of noise (with inner indices taking care of the noise simulation) and quantum entanglement (with bond indices taking care of two-qubit gate simulation). Consequently, in case of weak system noise, the resource cost of MPDO will be significantly less than that of the MPO due to a relatively small inner dimension needed for the simulation. In case of strong system noise, a relatively small bond dimension may be sufficient to simulate the noisy circuits, indicating a regime that the noise is large enough for an `easy' classical simulation, which is further supported by a comparison with the experimental results on an IBM cloud device. Moreover, we propose a more effective tensor updates scheme with optimal truncations for both the inner and the bond dimensions, performed after each layer of the circuit, which enjoys a canonical form of the MPDO for improving simulation accuracy. With truncated inner dimension to a maximum value $\kappa$ and bond dimension to a maximum value $\chi$, the cost of our simulation scales as $\sim ND\kappa^3\chi^3$, for an $N$-qubit circuit with depth $D$.
\end{abstract}

\maketitle
\renewcommand\theequation{\arabic{section}.\arabic{equation}}
\setcounter{tocdepth}{4}
\makeatletter
\@addtoreset{equation}{section}
\makeatother

\section{Introduction}
\label{sec:intro}
Quantum computer has the potential to outperform the best possible classical computers in many tasks such as factoring large numbers. It relies on the fact that wavefunctions represent amplitudes that grow exponentially in terms of the system size~\cite{nielsen2000quantum}. At the heart is quantum coherence, which is fragile and easily destroyed by noise. In principle, this drawback may be overcome by the techniques of quantum error correction and fault-tolerance, which however require tens of thousands of qubits to perform computing tasks of practical relevance~\cite{shor1995scheme,steane1996multiple}. For near-term hardware systems, precision needs to improve while systems size grows, to be able to perform reasonable computing tasks before decoherence, whose performance may be measured by the so-called quantum volume~\cite{bishop2017quantum}. It is claimed that systems with quantum volume as large as $32$ have been achieved~\cite{IBM32}, and systems of quantum volume $64$ is on the way~\cite{gaebler2019progress}.

Real world quantum computers battling noise recently achieved the so-called quantum supremacy, at Google, for implementing random quantum circuits in a $53$-qubit system and a circuit depth of $20$~\cite{arute2019quantum}. It is well-known that noiseless random circuits are hard to simulate on classical computers \cite{aaronson2005quantum,bremner2011classical,aaronson2011computational,
fujii2017commuting,bremner2016average,aaronson2016complexity}, and simulations of (noiseless) random quantum circuits on supercomputers have been implemented for more than $40$ qubits \cite{larose2019overview,de2019massively,smelyanskiy2016qhipster,
jones2019quest}. It still remains unclear, however, whether there are classical algorithms running on available supercomputers that may be able to simulate the behavior of the Google system, due to physical noise that results in low fidelity compared to noiseless systems. For instance, a method based on second storage has been proposed, which may simulate the system in a few days~\cite{pednault2019leveraging}. 

It is recently proposed in~\cite{zhou2020limits} that a method based on Matrix Product States (MPS) (one of the one-spacial-dimensional Tensor Network states) can approximate the behavior of real quantum systems. The MPS has been a powerful method that faithfully represents ground states of local Hamiltonians~\cite{Verstraete:2008ko,2011AnPhyS,Orus2014}. The Singular Value Decomposition (SVD) method for truncating the bond dimension for MPS has been shown great success for finding ground states of one-spacial-dimensional (1D) local Hamiltonians, both gapped and gapless. It is unclear, however, what is the error model that the MPS method represents, for simulating circuit output distribution of real quantum computers.

Since MPS cannot represent mixed states of quantum systems, a natural idea is instead to use the Matrix Product Operators (MPO)~\cite{Pirvu_2010,verstraete2004matrix}. The MPO method has been used to simulate quantum circuits of Shor's and Grover's algorithm with noise~\cite{woolfe2015matrix}. Very recently, the MPO method has also been used to simulate 1D random circuits~\cite{Noh2020efficient}. For simulating two-qubit gates, the MPO tensors with a 'canonical' update bring a factor of $DN^2$ in simulating an $N$-qubit random circuit with depth $D$ in terms of complexity.

In this work, we simulate noisy 1D random quantum circuits with Matrix Product Density Operators (MPDO), based on the MPDO construction proposed in~\cite{verstraete2004matrix}. Recently, it is also shown that 
MPDO can describe many-body thermal states efficiently~\cite{jarkovsky2020efficient}. Compared with the MPO method, the inner indices in the MPDO method capture the classical information of the noise simulation, which also reduces the computational and memory complexity under the condition of weak noise. The MPDO model consists of two parts that are conjugated to each other. By so, the simulation of the two-qubit gates can be done in a similar way as the MPS simulation, which is taken care of by the bond indices. In case of weak system noise, a small inner dimension may be sufficient for the simulation, so the resource cost of MPDO could be significantly less than that of the MPO. In case of strong system noise, a relatively small bond dimension may be sufficient to simulate the noisy circuits, indicating a regime that the noise is large enough for an `easy' classical simulation. 

Moreover, we propose a more effective canonical tensor update scheme, performed after each layer of the circuit, which would truncate the inner dimension to some maximum value $\kappa$ and the bond dimension to some maximum value $\chi$ with a canonicalization of the MPDO for improving simulation accuracy. The complexity of this scheme only proportional to $DN$ for an $N$-qubit circuit with depth $D$. The cost of our entire simulation scales as $\sim ND\kappa^3\chi^3$. 

We apply our method to simulate the random quantum circuit with different noise models, including the dephasing noise, the depolarizing noise, and the amplitude damping noise. We demonstrate that MPDO approximates the noisy output quantum states well, while the method based on Matrix Product States (MPS) fails to approximate the noisy output quantum states for any of the noise models considered. This indicates that the bond dimension truncation method of the MPS simulation might not represent any local noise model in real physical systems. With a further look into the deviation from the Porter-Thomas distribution for the ideal random circuit case, relatively small bond dimension for the MPDO method already grasp some `qualitative behavior' of the noisy output distribution. To test our method with a real quantum computer, we run random circuits on an IBM $16$-qubit device. The comparison between experimental data and the simulation based on MPDO method demonstrates that relatively small $\chi$ and $\kappa$ can indeed simulate the noisy random circuit efficiently. 

We organize our paper as follows. In Sec.~\ref{sec:noise}, we discuss the error models we use for our circuit simulation. In Sec.~\ref{sec:MPDO}, we discuss the MPDO method for simulating noisy quantum circuits and its complexity. In Sec.~\ref{sec:comparison}, we present our results based on the MPDO method, and compare with an exact noise simulation based on density matrices, and the MPS method based on bond dimension truncation, for different error models. In Sec.~\ref{sec:truncation}, we study the effect of truncation on the bond and inner dimensions. In Sec.~\ref{sec:IBM}, we run several 1D random circuits on $10$ qubits of the $16$-qubit IBM device ibmq$\_$16$\_$melbourne, and compare with our simulation. In Sec.~\ref{sec:QEC}, we apply our MPDO method for simulating noisy encoding circuits of a quantum error-correcting code. Summary of the results and discussions on future directions will be given in Sec.~\ref{sec:discussion}.

\section{Noise models}
\label{sec:noise}
Physical noise $\cal{E}$ for quantum systems with a quantum state $\rho$ are generally characterized by the Kraus representation, as given by
\begin{equation}
{\cal{E}}(\rho)=\sum_k E_k\rho E_k^{\dag},
\label{eq:kraus}
\end{equation}
where $E_k$ s are the Kraus operators and fulfill $\sum_k E_k^{\dag} E_k=I$~\cite{nielsen2000quantum}.

For the quantum circuit of $N$ qubits with single-qubit and two-qubit quantum gates, normally the fidelity of single-qubit gates are much higher than that of the two-qubit gates. We will then assume that all single-qubit gates are ideal, and model the noise only on the two-qubit gate $U$ by
\begin{equation}
\label{eq:Kraus}
\rho\rightarrow \sum_k UE_k \rho E_k^{\dag}U^{\dag} ,
\end{equation}
where $E_k$s acting on the same qubits of $U$.

Alternatively, for the noisy channel by $\mathcal{E}$, we can denote the channel corresponding to the two-qubit gate $U$ by $\mathcal{U}$, hence re-write Eq.~\eqref{eq:Kraus} as
\begin{equation}
\rho\rightarrow\mathcal{U}\circ\mathcal{E} (\rho).
\end{equation}

In this work, we consider the following noise models.

\begin{itemize}

\item Dephasing noise

The dephasing noise on a single qubit can be modeled by
\begin{equation}
\rho\rightarrow \mathcal{E}_{DF}(\rho)=(1-\epsilon) \rho + \epsilon Z\rho Z^{\dag},
\end{equation}
where $\epsilon \in [0, 1]$, $Z$ is the Pauli operator.

For a noisy two-qubit gate $U$, we model the dephasing noise by 
\begin{equation}\label{eq:dephasing}
\rho\rightarrow\mathcal{U}\circ\mathcal{E}_{DF}^{\otimes 2} (\rho).
\end{equation}

\item Depolarizing noise

The depolarizing noise on a single qubit can be modeled by
\begin{equation}\label{eq:depolarizing}
\rho\rightarrow \mathcal{E}_{DP}(\rho)=(1-\epsilon) U\rho U^{\dagger} + \epsilon \frac{I}{2}.
\end{equation}

For a noisy two-qubit gate $U$, we model the dephasing noise by 
\begin{equation}\label{eq:dephasing}
\rho\rightarrow\mathcal{U}\circ\mathcal{E}_{DP}^{\otimes 2} (\rho).
\end{equation}

Notice that, for a quantum circuit with gate noise modelled by depolarizing noise, the density matrix of the output state under will also be given in the form of a global depolarizing noise, i.e. 
\begin{equation}
\label{eq:depolarizing0}
\rho\rightarrow(1-\alpha)\ket{\psi}\bra{\psi}+\alpha \frac{I}{M},
\end{equation}
where $\ket{\psi}$ is the corresponding noiseless output state,
and $M=2^N$ with $N$ the number of qubits in the system.

\item Amplitude damping noise

The amplitude damping noise on a single qubit can be modeled by
\begin{equation}
\rho\rightarrow \mathcal{E}_{AD} (\rho)=A_0 \rho A_0^{\dag}+  A_1 \rho  A_1^{\dag},
\label{eq:adn}
\end{equation}
where 
$A_0=\begin{pmatrix}
1 & 0 \\
0 & \sqrt{1-\epsilon} 
\end{pmatrix},\ 
A_1=\begin{pmatrix}
0 & \sqrt{\epsilon}  \\
0 & 0
\end{pmatrix}.$

For a noisy two-qubit gate $U$, we model the amplitude damping noise by 
\begin{equation}\label{eq:damping}
\rho\rightarrow\mathcal{U}\circ\mathcal{E}_{AD}^{\otimes 2} (\rho).
\end{equation}

\end{itemize}

\section{Modeling noise simulation by Matrix Product Density Operators (MPDO)}
\label{sec:MPDO}
\begin{figure}[hbt]
\centering

  \includegraphics[scale=0.2]{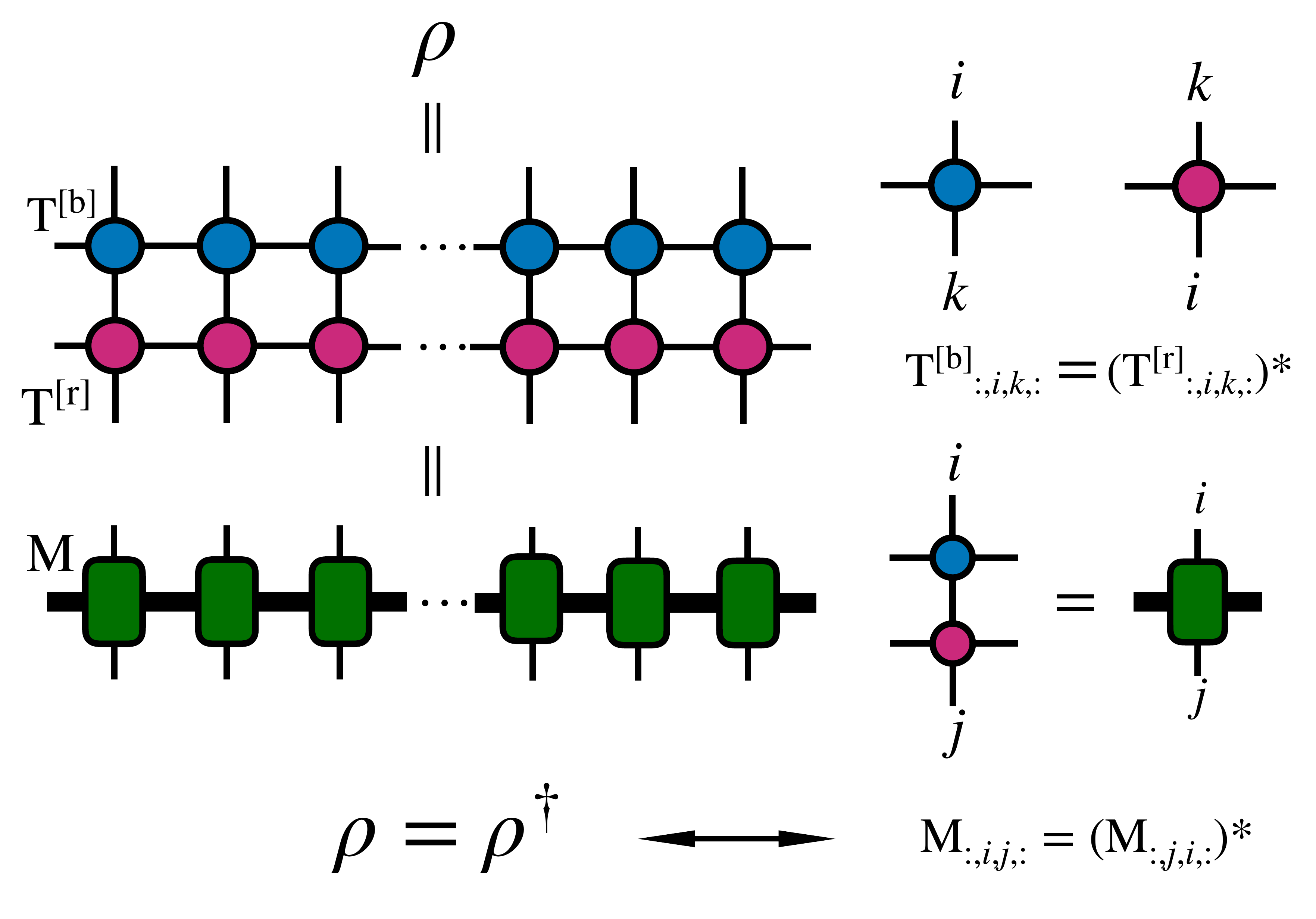}
  \caption{The structure of the Matrix Product Density Operators (MPDO). Instead of directly using a tensor $M$ to form a Matrix Product Operator, we choose $M$ to be composed by the red tensor $T^{[r]}$ and the blue tensor $T^{[b]}$, where $T^{[b]} = (T^{[r]})^\dagger$. By so, the MPDO will automatically satisfy the hermiticity of a density matrix, although $T$s are general $4$th order tensors. In this way, the parameters of the model can be fully used on the one hand, and the inner indices between the $T^{[b]}$ and $T^{[r]}$ have emerged on the other hand. Those inner indices can be used to describe the classic entropy of the system. At the same time, when the classic entropy is small, the model's computational cost will become significantly smaller.}
  \label{fig:structure}
\end{figure}
\begin{figure}

  \includegraphics[scale=0.2]{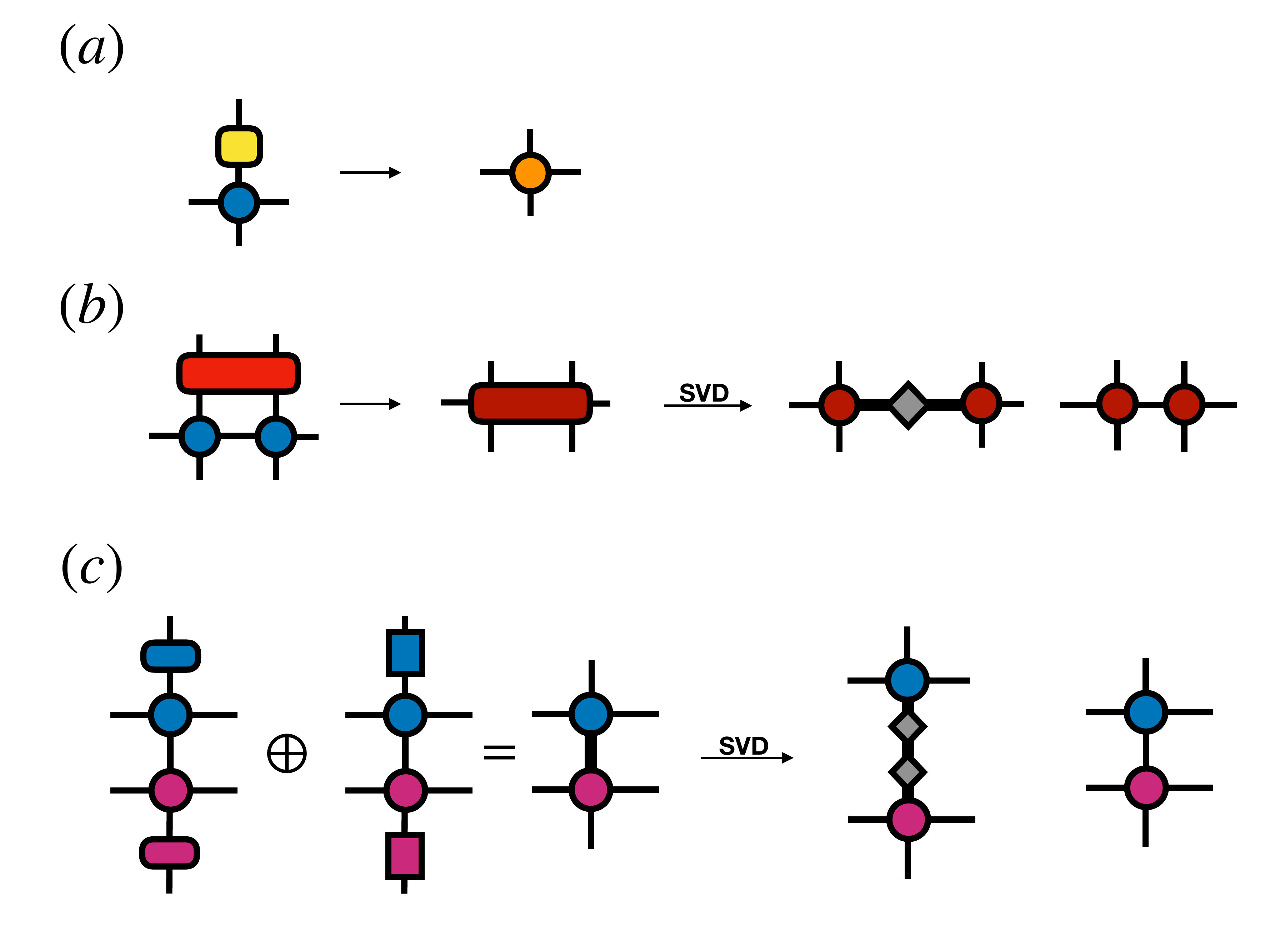}
  \caption{\textbf{(a)} Applying a single-qubit gate on the corresponding tensor of the MPDO. \textbf{(b)} Applying a two-qubit gate on two corresponding tensors of the MPDO. The new tensors would be obtained by the singular value decomposition (SVD). The SVD would increase the bond dimension between those two tensors, while we could find the global optimal truncation of tensors based on the singular value of SVD. \textbf{(c)} Applying noise models on the corresponding tensor is equivalent to applying each noise operator on tensor separately and then direct sum them on the inner indices. The increased dimensions of inner indices can also be truncated using SVD.}
  \label{fig:gate_noise}
\end{figure}

By applying the Matrix Product Operators(MPO) instead of the Matrix Product States(MPS), one can extended the error model to represent mixed quantum states, which allows us to introduce typical random noise in a more direct and efficient way. As Fig. \ref{fig:structure} shown, we use the MPO to represent the density matrix $\rho$ of N qubits
\begin{equation}
\begin{split}
\rho = \sum_{s_1,s'_1...,s_N,s'_N=1}^{2}\Tr(M_{L_1,R_1}^{s_1,s'_1}M_{L_2,R_2}^{s_2,s'_2}  \cdots M_{L_N,R_N}^{s_N,s'_N}) 
\\ \times  \ketbra{s_1, \cdots, s_N}{s'_1, \cdots, s'_N}
\label{eq:mpdo}
\end{split}
\end{equation}
where $\rho$ denotes the density matrix of a mixed quantum state. $s_k$ and $s'_k$ are known as the "physical" indices, which expand into the $2^N \times 2^N$ density matrix $\rho$. $L_k$ and $R_k$ denote the left and right "bond" indices, which carry some information of entanglement between qubits. 

However, there are at least two reasons that directly using the tensor $M$ to form an MPO representation of the density matrix may not be the best choice. First, since the density matrix is a Hermitian matrix, we must guarantee that $M = M^\dagger$, which will make at least half of the M parameters invalid, resulting in additional computational overhead. Second, considering the density matrix $\rho = \sum_k \lambda_k \ketbra{\phi_k}{\phi_k}$, when $k$ is not too many, the density matrix can be regarded as the sum of the direct products of several state vectors. In this case, it is very inefficient to use tensor networks to model the direct product of state vectors instead of the sum of several states itself. 

Therefore, a proper way to ensure the hermiticity of density matrix $\rho$ is the Matrix Product Density Operators(MPDO)~\cite{PhysRevLett.93.207204}, which design the $M_{L_k,R_k}^{s_k,s'_k}$ to be composed by a general 4th order tensor $T_{l_k,r_k}^{s_k,a_k}$ and its conjugate copy $(T_{l'_k,r'_k}^{s'_k,a_k})^\ast$. 
\begin{equation}
M_{L_k,R_k}^{s_k,s'_k} = \sum_{a_k=1}^{d^k} T_{l_k,r_k}^{s_k,a_k} \times (T_{l'_k,r'_k}^{s'_k,a_k})^\ast.
\label{eq:conjugate_tensor}
\end{equation}
In others words, the MPDO is composed of a general Matrix Product Operators(MPO) and its conjugation. The indices $L_k$ and $R_k$ of $M$ correspond to the direct product of indices $l_k \otimes l'_k$ and $r_k \otimes r'_k$ respectively. In practice, the dimension of $l_k$ and $r_k$ bond indices $D_k$ are restricted to a certain maximum value $\chi$.  

The two conjugate MPOs were connected by the "inner" indices $a_k$, which could carry the classical information of the system. This `inner' dimensions will increase by adding statistical noise. The dimension of the inner indices $d_{k}$ would be no more than $2D_kD_{k+1}$. While in some cases, we still truncate the $d_k$ to a smaller number $\kappa < 2D_kD_{k+1}$. The memory cost of the MPDO would only proportional to $2N\kappa\chi^2$. Note that if we use the $M$ matrix directly, the memory cost will be $4N\chi^4$, which means that if $\kappa$ reaches the upper bound, those two models cost almost the same memory. While if the system noise is small, the dimensions of the inner indices will be fixed at a smaller value, and the model will consume significantly fewer resources than using $M$ directly.

To be more intuitive, here we give two particular examples of the MPDO. The first example is the density matrix of a pure quantum state $\rho_{\mathrm{pure}} = \ketbra{\mathbf{s}}{\mathbf{s}}$, where $s_k \in \{0, 1\}$. The corresponding $\mathrm{T}^{(k)}$ is $\{\mathrm{T}_{l_k,r_k}^{s_k,a_k}: T^{s_{k},0}_{0,0} = 1,\mathrm{others}: 0\}$. For the density matrix of a maximum mixed state $\rho_{\mathrm{mix}} = (\frac{1}{2}\ketbra{0}{0} + \frac{1}{2}\ketbra{1}{1})^{\otimes N}$, the $\mathrm{T}^{(k)}$ can be written as $\{\mathrm{T}_{l_k,r_k}^{s_k,a_k}: T^{0,0}_{0,0} = T^{1,1}_{0,0} = \frac{1}{\sqrt{2}}, \mathrm{others}: 0\}$.

Applying gates and noise on the MPDO is straightforward. Considering the symmetry of the density matrix, one only needs to apply the gate to half of the density matrix, which is the MPO, the rest automatically becomes the conjugation of updated tensors. It is worth mentioning that the identity $I$ of the depolarizing noise in (\ref{eq:depolarizing}) can also be decomposed into two conjugate parts using $2I = \rho + \sum_{i = x,y,z}\sigma_i\rho \sigma_i^{\dag}$, where $\sigma_i$ are Pauli matrices. As Fig. \ref{fig:gate_noise} (a) shown, the one-qubit gate wouldn't change any topology or dimensions of the MPDO, we could do it exactly with complexity $\sim\kappa \chi^2$.

A two-qubit gate would be represented as a $4$th order tensor $U$. As shown in Fig.~\ref{fig:gate_noise} (b), we first contract the gate with the corresponding qubits, which form a $6$th order tensor $W$. It can be separated into two new MPO tensors by the Singular Value Decomposition (SVD). In general cases, applying a two-qubit gate would increase the dimension of bond indices between the two qubits to $\mathrm{min}(2D_kd_k, 2D_{k+2}d_{k+2})$. If we truncate it to $\chi$, the approximation error is $\sim \sum_{j = \chi + 1}{\lambda^2_j} / \sum_{i=1}\lambda^2_i $, where $\lambda_i$ are the singular values in descending order. More details about the strategy of this truncation are placed at the end of this section. The computation cost is $\sim \kappa^2\chi^3$ for contraction, and $\sim \kappa^3\chi^3$ for SVD. 

Compared with the MPS method as discussed in~\cite{zhou2020limits}, the MPDO has a clear advantage in adding noise. Unlike the state representation, the density matrix can directly express noise as the Kraus representation in (\ref{eq:kraus}), which avoids repeated Monte Carlo sampling of different noise. Let's take the amplitude damping noise as an example. What we need to do is to apply the $A_0$ and $A_1$ in (\ref{eq:adn}) to $\rho$ respectively and then add them directly. Same to the gate operators, we only need to apply noise to half of $\rho$ to save the computational cost. Note that the summation of noise and the contraction of the conjugate tensors are not interchangeable. To avoid unnecessary cross-terms, as shown in Fig.~\ref{fig:gate_noise}~(c), here we need direct sum the different noise parts on the inner indices $a_k$. which writes,
 \begin{equation}
 T_{l_k,r_k}^{s_k,a_k} =  
 \begin{cases}
 \sum_{s'_k}A_0^{s_k, s'_k}T_{l_k,r_k}^{s'_k,a'_k} & a_k \in [0, d_k)\\
  \sum_{s'_k}A_1^{s_k, s'_k}T_{l_k,r_k}^{s'_k,a'_k} & a_k \in [d_k, 2d_k),
  \end{cases}
\end{equation}
thus, 
 \begin{equation}
M^{[\mathrm{new]}} = \sum_{a_k} T T^\dagger= A_0M^{[\mathrm{old}]}A^{\dagger}_0+ A_1M^{[\mathrm{old}]}A^{\dagger}_1.
\end{equation}
The computational cost of applying noise is $\sim m\kappa\chi^2$, $m$ is the number of terms of the noise model.

There is an interesting fact behind the structure of the MPDO. Note that the two-qubit gate introduces entanglement entropy into the system, and it only increases the dimension of the bond indices; meanwhile the single-qubit noise introduces classical statistical entropy into the system, and only increases the dimension of the inner indices. Therefore, if there is no truncation, in the final structure of the MPDO, the dimensions of the bond indices and inner indices will be related to the quantum entanglement entropy and classical statistical entropy of the system, respectively. The exact relationship between bond/inner dimensions and entanglement/classical entropy remains an open question. 

In many cases, especially those with significant noise, a small dimension of bond indices is enough to ensure high fidelity. 
So in order to remove unnecessary parameters, we may truncate the bond indices and inner indices,  simultaneously. More specifically, we will first do a local optimal approximation on inner indices by SVD, which separate tensor $T$ into
\begin{equation}
T_{l_k,r_k}^{s_k,a_k} = \sum_{\mu} = U_{l_k,r_k}^{s_k,\mu}S_{\mu}V_{\mu,a_k},
\end{equation}
where $S_{\mu}$ is the singular value of $T$, $U$ and $V$ are unitary matrix. Then we only keep $\kappa$ largest $S_{\mu}$ and corresponding orthogonal vectors in $U$ and $V$. The approximate $M$ can be write as
\begin{equation}
M \approx \sum^{\kappa}_{\mu=1} U'S^2U'^{\dagger}.
\end{equation}
We repeat this process on each inner indices. Then, before we truncate bond indices, we first apply QR decomposition from the left of MPDO to the right to form a canonical form of the MPO, which writes
\begin{equation}
\begin{split}
T_{l_k,r_k}^{s_k,a_k} &= \sum_{\mu}  Q_{l_k,\mu}^{s_k,a_k}R_{\mu,r_k}, \\
T_{l_k,r_k}^{'s_k,a_k} &= Q_{l_k,\mu}^{s_k,a_k}, \\
T_{l_{k+1},r_{k+1}}^{'s_{k+1},a_{k+1}} &= \sum_{l_{k+1}} R_{\mu,l_{k+1}}T_{l_{k+1},r_{k+1}}^{s_{k+1},a_{k+1}}.
\end{split}
\end{equation}
After applying this on all qubits, all tensors except the rightmost one are left canonicalized. In other words, they fulfill
\begin{equation}
\sum_{l_k,s_k,a_k}T_{l_k,r_k}^{s_k,a_k}T_{l_k,r'_k}^{s_k,a_k} = \mathrm{I}_{r_k,r'_k},
\end{equation}
where $\mathrm{I}$ denotes the identity matrix. This canonicalization will ensure the SVD truncation on rightmost tensor is globally optimal. We then use the SVD to truncate each bond indices from right to left. 
\begin{equation}
\sum_{l_{k+1}}T_{l_k,l_{k+1}}^{s_k,a_k}T_{l_{k+1},r_{k+1}}^{s_{k+1},a_{k+1}}\approx \sum^{\chi}_{\mu = 1}U_{l_k,\mu}^{s_k,a_k}S_{\mu}V_{\mu,r_{k+1}}^{s_{k+1},a_{k+1}}.
\end{equation}
Note that each time the SVD changes the right tensor from left canonicalized to right canonicalized, which results in the following SVD truncations are all globally optimal. Therefore, the most economical way is first to complete a layer of two-qubit gates and noise (see Fig.~\ref{fig:1DR} for an illustration of a layer in the dotted line circuit), then to perform a canonicalization from left to right and the following SVD decomposition from right to left. Compared with canonicalizing on each qubit independently, the order of this scheme reduces the complexity of canonical truncation from a factor of $N^2$ to a factor of $N$ without loss of the accuracy. The total cost of simulating an N-qubit circuit with depth $D$ is $\sim DN\kappa^3\chi^3$. Note that while the canonical form of the MPO can not be used to find the global optimal truncation of inner indices. we could still use the full update method similar to that used in the higher dimensional tensor network to find its global optimal truncation, in case there are some people be willing to tolerate excessive calculation costs. 

\section{Comparison MPDO simulation with different models}
\label{sec:comparison}

\begin{figure*}[htbp]
	\centering
	\subfigure[Dephasing]{
		\begin{minipage}[t]{0.33\linewidth}
			\centering
			\includegraphics[scale=0.35]{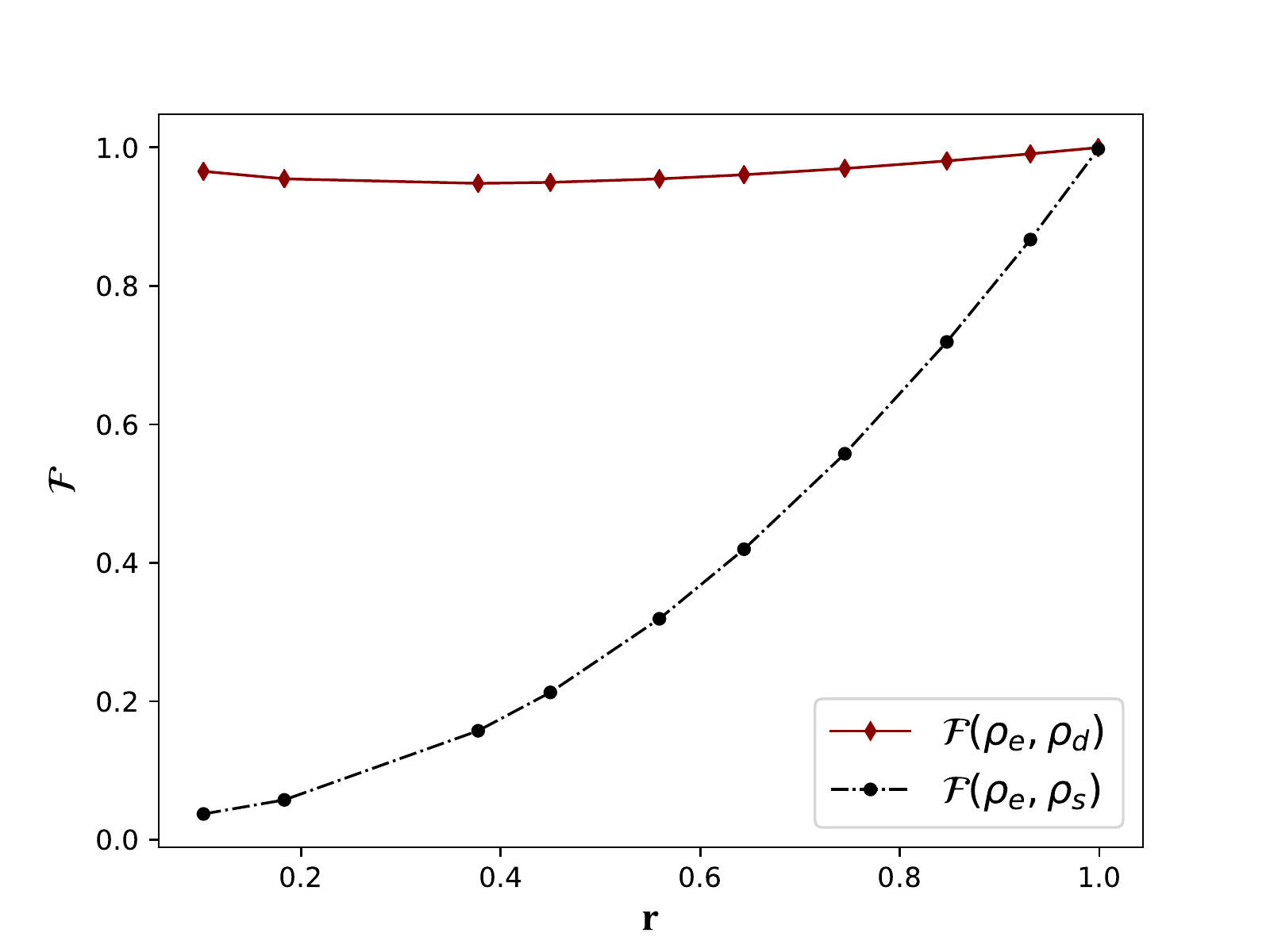}
		\end{minipage}%
	}%
	\subfigure[Depolarizing]{
		\begin{minipage}[t]{0.33\linewidth}
			\centering
			\includegraphics[scale=0.35]{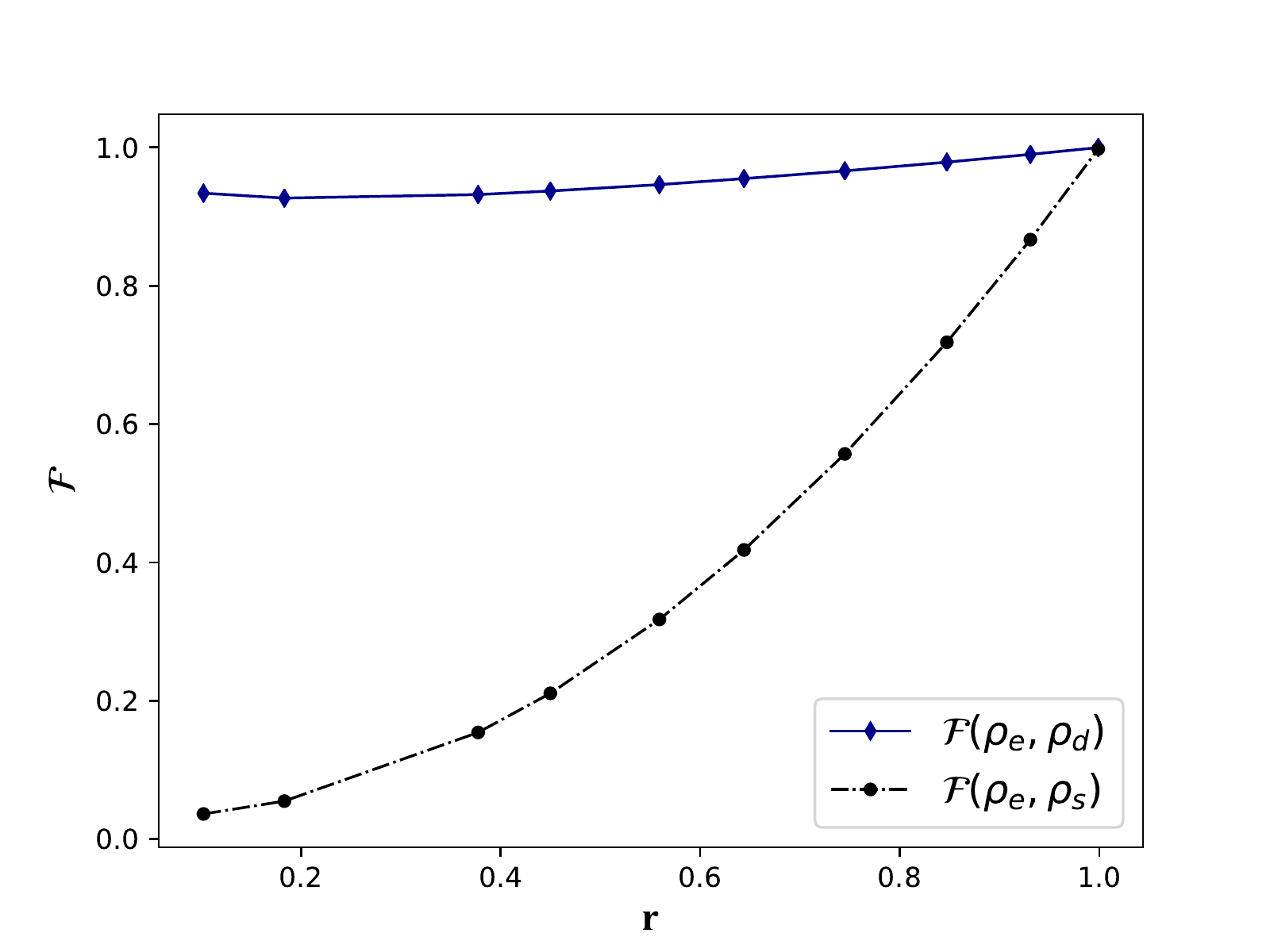}
		\end{minipage}%
	}%
	\subfigure[Amplitude damping]{
		\begin{minipage}[t]{0.33\linewidth}
			\centering
			\includegraphics[scale=0.35]{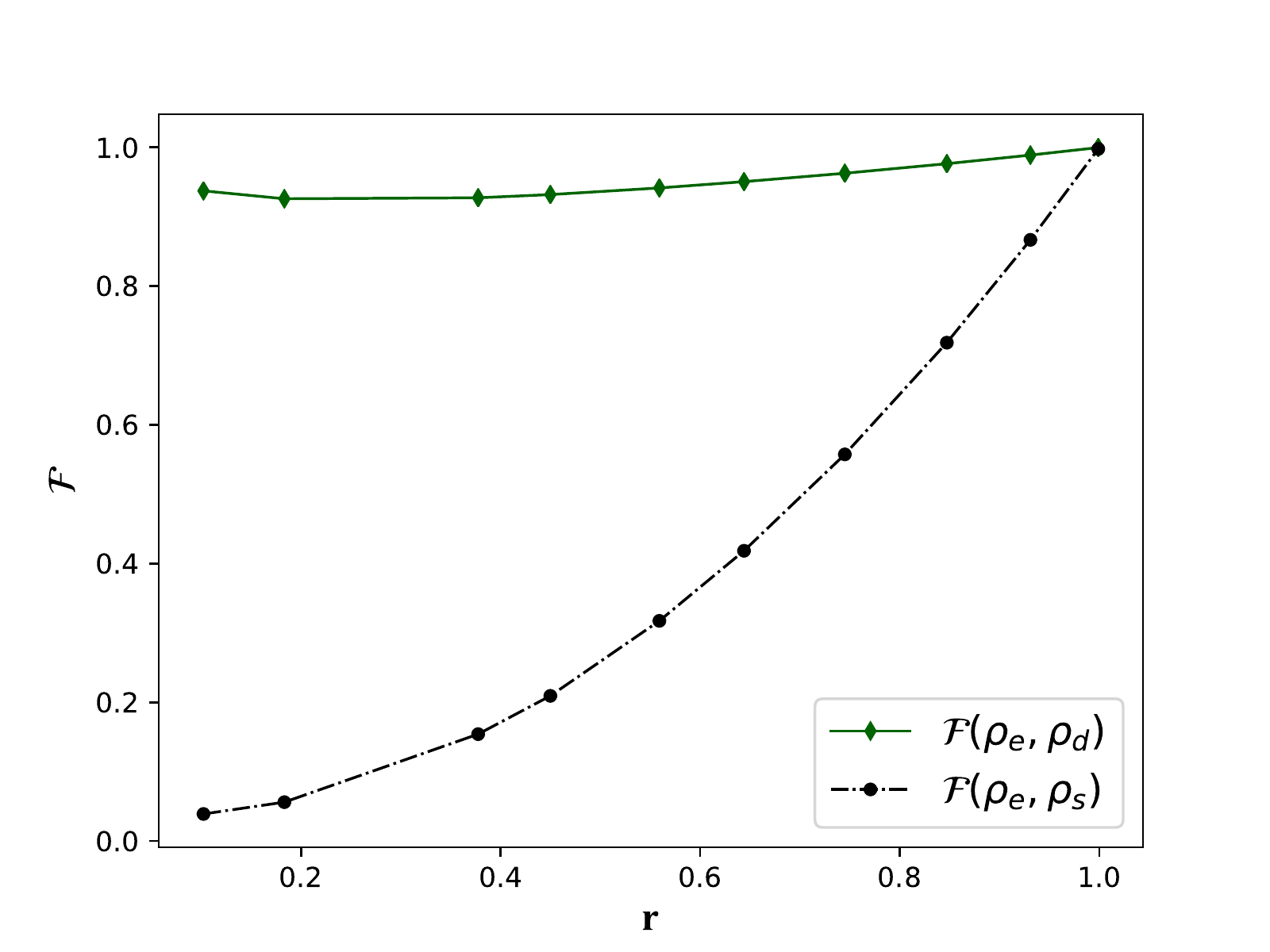}
		\end{minipage}
	}%
	\centering
	\caption{Fidelity comparison for $\rho_0$, $\rho_e$, $\rho_{d}$, and $\rho_{s}$, where $r=\mathcal{F}(\rho_{0}, \rho_{s})=\mathcal{F}(\rho_{0}, \rho_{e})$. (a) the dephasing noise; (b) the depolarizing noise; (c) the amplitude damping noise.}
	\label{fig:mps_mpo}
\end{figure*}

\begin{figure}[hbtp]
\centering
    \includegraphics[scale=0.4]{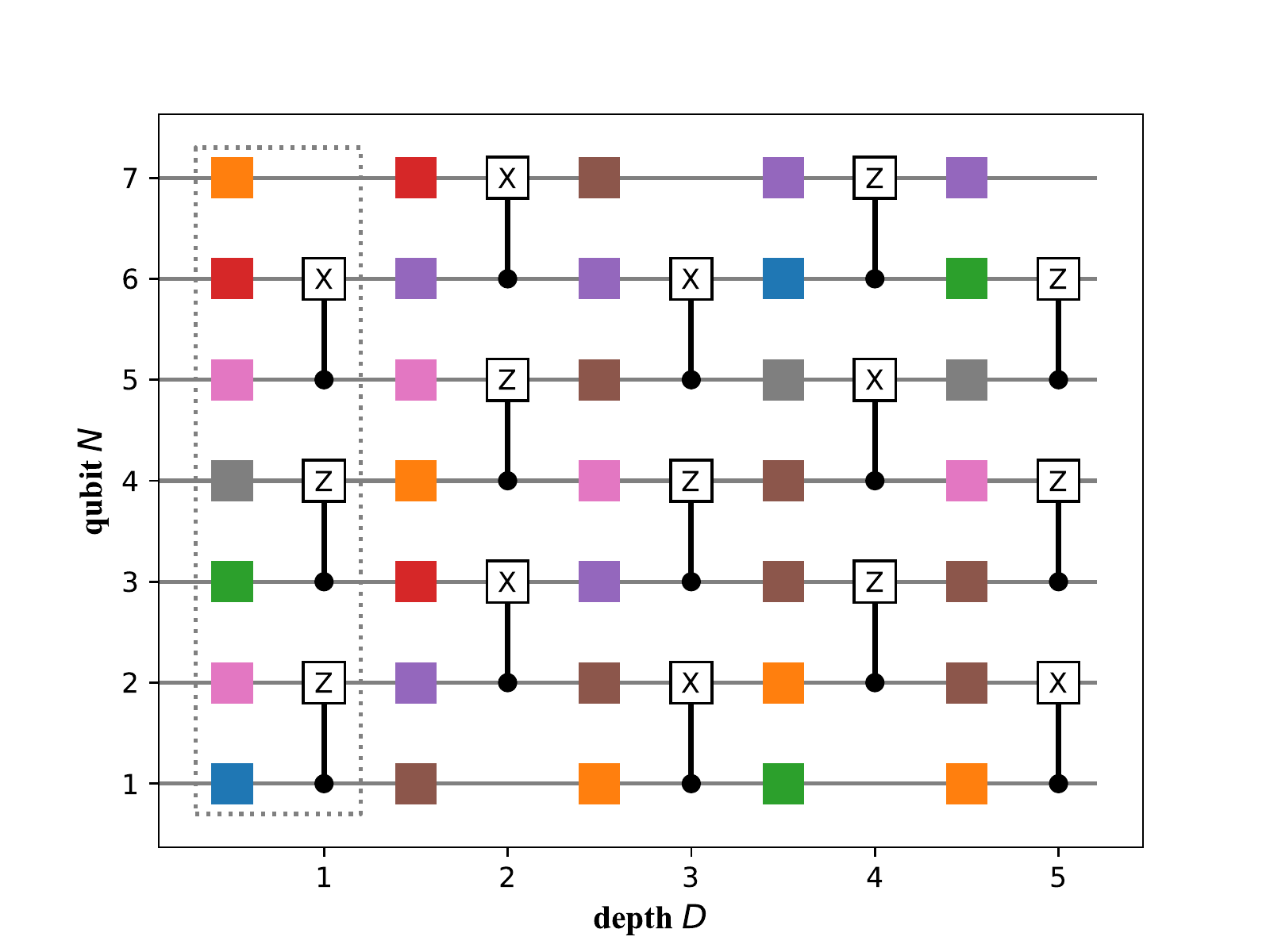}
    \caption{Sketch of the 1D random circuit with qubits number $N=7$ and depth $D=5$. The gray dotted line outlines the structure of one layer. The circuit is interleaved by layers of single-qubit gates (colored boxes) and two-qubit gates (the dots connected to a blank boxes). The single-qubit gates are randomly sampled from the set of universal single-qubit quantum gates. The two-qubit gates are either control-NOT or control-Z with equal probability.}
    \label{fig:1DR}
\end{figure}

Our numerical experiments are applied on the 1D random circuit illustrated in Fig.~\ref{fig:1DR}. The colored boxes represent various single-qubit gates randomly generated from $e^{i \alpha( \sigma _x\sin \theta \cos \phi +\sigma _y\sin \theta \sin \phi +\sigma _z\cos \theta )}$ with three random parameters $\alpha,\theta,\phi \in [0,2\pi)$, which traverse the space of universal single qubit gates. The lines and the blank boxes connecting two qubits represent either CNOT or Control-Z gates with equal probability. It is known that such kinds of (pseudo-)random circuit with big enough depth $D$ could yield an approximately Haar-distributed unitary, and generate entanglement efficiently~\cite{emerson2003pseudo,oliveira2007generic,harrow2009random}. 
This kind of (pseudo-)random quantum circuits has been discussed extensively for demonstrating quantum supremacy~\cite{aaronson2016complexity,boixo2018characterizing,neill2018blueprint,arute2019quantum}.

We perform four different models to simulate this circuit,

\begin{itemize}
\item A simulator based on the state vector for exact noiseless simulation.
\item A simulator based on the density matrix for exact noise simulation. The noise would apply to each two-qubit gate.
\item An MPS simulator based on approximating a pure state by SVD method, as discussed in~Ref.~\cite{zhou2020limits}.
\item An MPDO simulator based on approximating a density matrix by an conjugated tensor network structure and SVD, as discussed in Sec.~\ref{sec:MPDO}. The noise would apply to each two-qubit gate.
\end{itemize}

Various noise models has been considered, including the dephasing, the depolarizing, and the amplitude damping noise model, as discussed in Sec.~\ref{sec:noise}.

\subsection{Comparison based on fidelity}
\label{sec:fidelity}

We consider the random circuits with $10$ qubits and depth $D=24$. 
For clarity, we define the following notations.
\begin{itemize}
\item $\rho_0$ denotes the output density matrix of the exact noiseless simulator, which corresponds to a pure state representing the exact output the noiseless random circuit.
\item $\rho_e$ denotes the output density matrix of the exact noise simulator, which gives the exact result of simulating the circuit with given noise models.
\item $\rho_s$ denotes the output density matrix of the MPS simulator, which corresponds to a pure state subject to different truncation up to some maximum bond dimension $\chi$.
\item $\rho_d$ denotes the output density matrix of the MPDO simulator, which is subject to different truncation up to some maximum bond dimension $\chi$ and maximum inner dimension $\kappa$.

\end{itemize}

The fidelity between two quantum states are given by
\begin{equation}
\mathcal{F}\left(\rho, \sigma\right) \equiv \operatorname{Tr} \sqrt{\sqrt{\rho} \sigma \sqrt{\rho}}.
\end{equation}

For comparing the MPS simulator with others, we define a parameter $r$ as
\begin{equation}
r=\mathcal{F}(\rho_{0}, \rho_{s}),
\end{equation}
which corresponds to a certain bond dimension truncation $\chi$. 
For each $\chi$ and noise model, we could find a corresponding error rate $\epsilon$ in $\rho_e$ that satisfied
\begin{equation}
\mathcal{F}(\rho_{0}, \rho_{e}) = r.
\end{equation}
We summarize the values of $\chi$ and corresponding $\epsilon$ in Table~\ref{tab:noise_rates}.

\begin{table}[htbp]
	\centering
	\begin{tabular}{ |p{1.1cm}||p{0.9cm}|p{1.9cm}|p{2cm}|p{1.9cm}|  }
		\hline
		Fidelity with $\rho_{0}$ & MPS bond dim & Dephasing noise rate & Depolarizing noise rate & Amplitude damping noise rate\\
		\hline
		0.102   & 2  & 0.0231 & 0.0302 & 0.0454\\
		0.183   & 3 & 0.0167 & 0.0220 & 0.0332\\
		0.378   & 4 & $9.47 \times 10^{-3} $& 0.0125 & 0.0188\\
		0.450  & 5 & $7.75 \times 10^{-3}$ & 0.0102 & 0.0155\\
		0.559   & 6 & $5.63 \times 10^{-3}$ &$7.45 \times 10^{-3} $& 0.0113\\
		0.644   & 7 & $4.25 \times 10^{-3}$ &$5.64 \times 10^{-3} $&$ 8.51 \times 10^{-3}$\\
		0.745   & 9 &$2.84 \times 10^{-3} $&$ 3.76 \times 10^{-3} $& $5.69 \times 10^{-3}$\\
		0.847   & 12 & $1.59 \times 10^{-3} $&$ 2.12 \times 10^{-3} $&$ 3.20 \times 10^{-3}$\\
		0.931   & 15 & $6.88 \times 10^{-4} $&$ 9.14 \times 10^{-4} $&$ 1.38 \times 10^{-3}$\\
		0.999   & 28 & $9.94 \times 10^{-6} $&$ 1.33 \times 10^{-5} $&$ 2.01 \times 10^{-5}$\\
		
		\hline
	\end{tabular}\\
	\caption{Noise rates in different noise models}
	\label{tab:noise_rates}
\end{table}

We then use the MPDO simulator to simulate the noisy random quantum circuit for different error models with error rate $\epsilon$ as given in Table~\ref{tab:noise_rates}. We set max bond dimension $\chi= 32$, and max inner dimension $\kappa= 48$.

For each $r$ and each noise model, we calculate the fidelity of $\mathcal{F}(\rho_{e}, \rho_{s})$, which demonstrates how the MPS method approximates the exact result of the noisy output density matrix, given the same fidelity of $r=\mathcal{F}(\rho_{0}, \rho_{s})=\mathcal{F}(\rho_{0}, \rho_{e})$. We also calculate the fidelity of $\mathcal{F}(\rho_{e}, \rho_{d})$, which demonstrates how the MPDO method approximates the exact result of the noisy output density matrix.
Our results are shown in Fig.~\ref{fig:mps_mpo}.

From Fig.~\ref{fig:mps_mpo}, it is clearly shown that when noise gradually became significant, the MPS simulator gradually failed to simulate all three noise models considered, even if it gives the fidelity with the exact noiseless state with $\rho_e$. This indicates that the MPS truncated approximation is not simulating physical noise in real systems. On the other hand, the MPDO method approximates $\rho_e$ well, which can, in fact, simulate any physical noise as given by the MPDO model construction. 

\subsection{Deviation from the Porter-Thomas distribution}
\label{sec:PT}

\begin{figure*}[htbp]
	\centering
	\subfigure[Dephasing]{
		\begin{minipage}[t]{0.33\linewidth}
			\centering
			\includegraphics[scale=0.35]{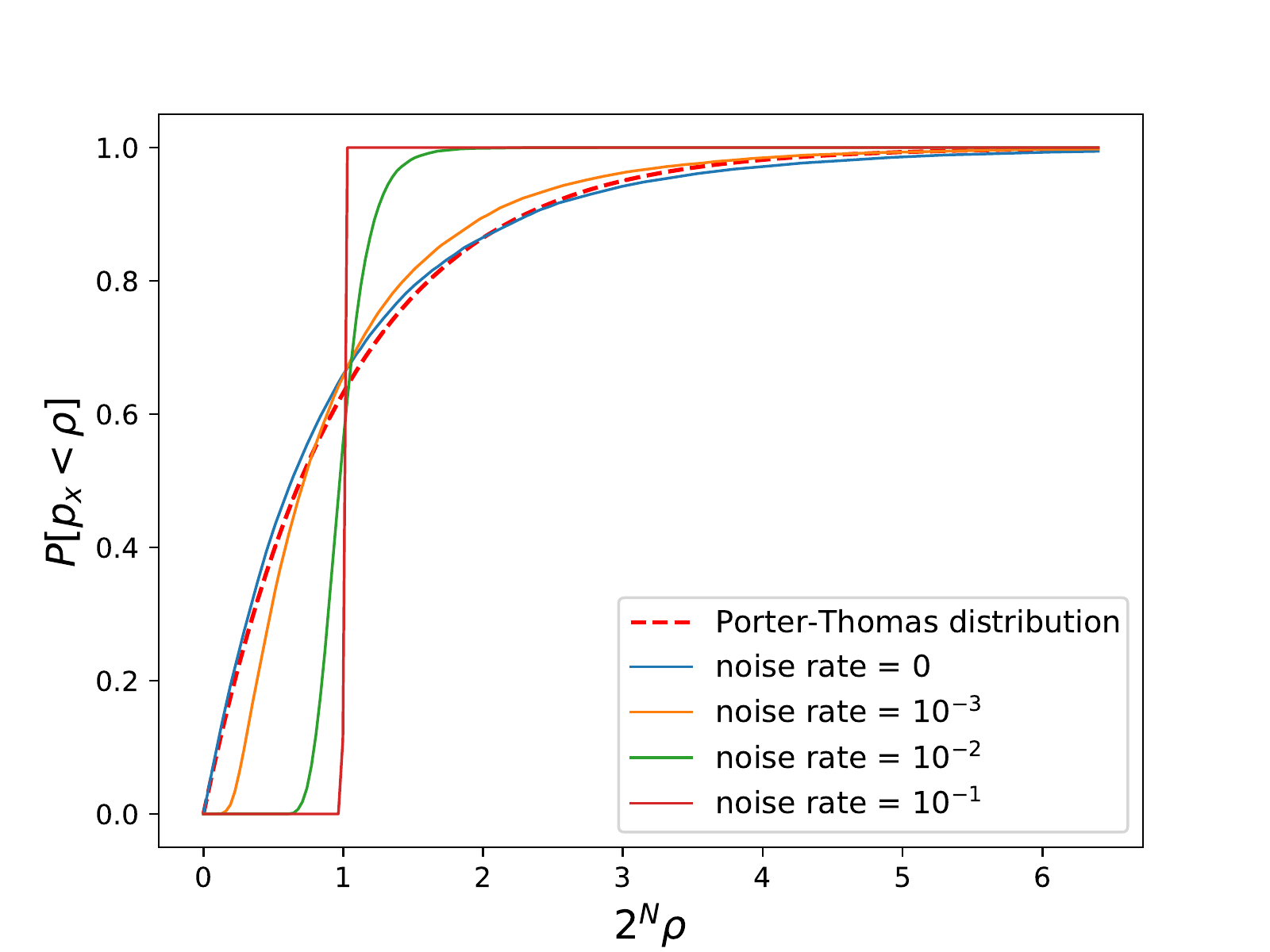}
		\end{minipage}%
	}%
	\subfigure[Depolarizing]{
		\begin{minipage}[t]{0.33\linewidth}
			\centering
			\includegraphics[scale=0.35]{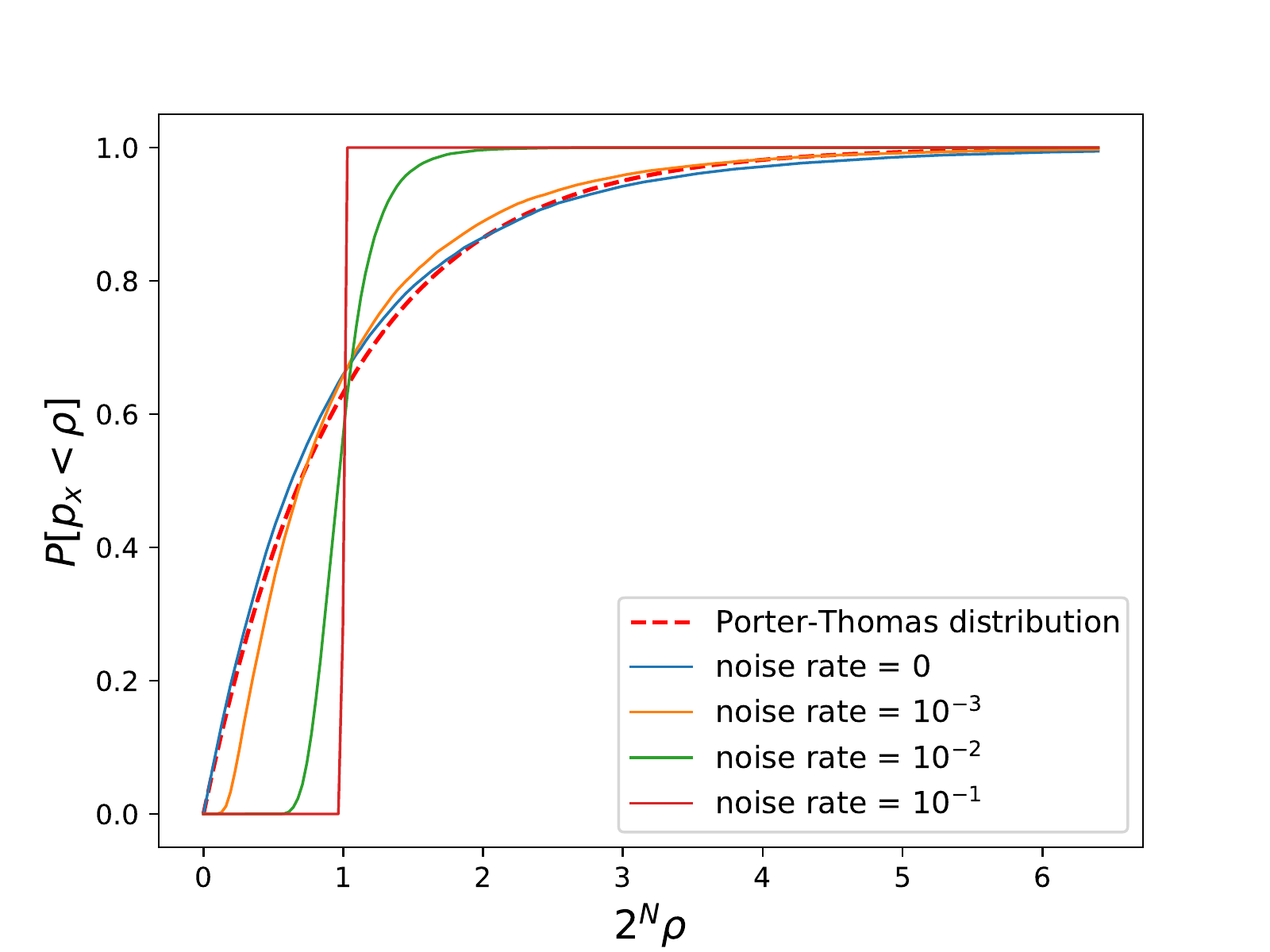}
		\end{minipage}%
	}%
	\subfigure[Amplitude damping]{
		\begin{minipage}[t]{0.33\linewidth}
			\centering
			\includegraphics[scale=0.35]{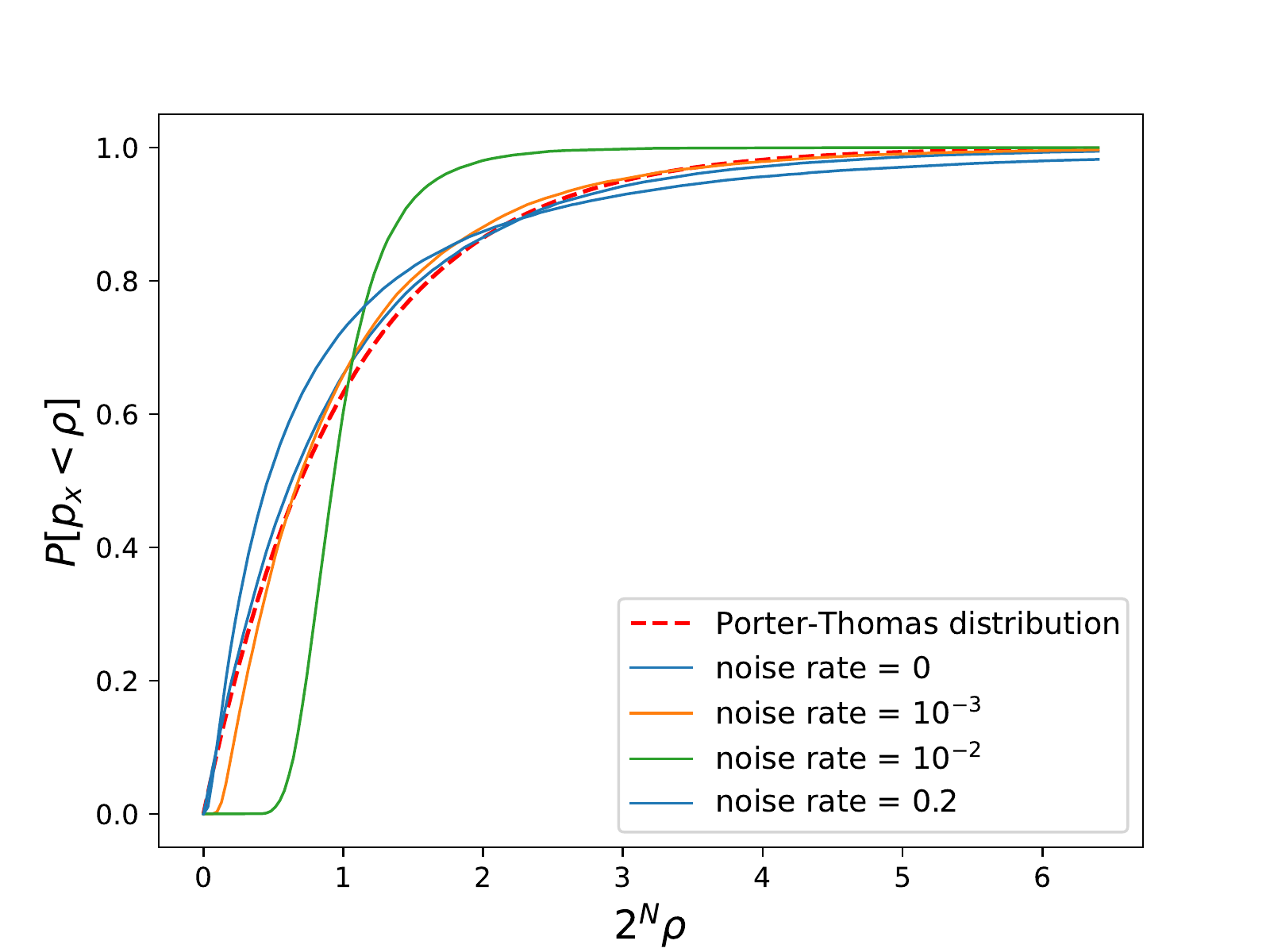}
		\end{minipage}
	}%
	\centering
	\caption{Exact noisy simulation of cumulated $p$ distributions for 1D random circuit with $15$ qubits and $24$ layers, under three different error models with different noise rate. 
	(a) the dephasing noise; (b) the depolarizing noise; (c) the amplitude damping noise. As the system noise increases, the cumulated $p$ distribution gradually deviates from the Porter-Thomas distribution.}
	\label{fig:pt_exact}
\end{figure*}

\begin{figure*}[htbp]
	\centering
	\subfigure[Dephasing]{
		\begin{minipage}[t]{0.33\linewidth}
			\centering
			\includegraphics[scale=0.35]{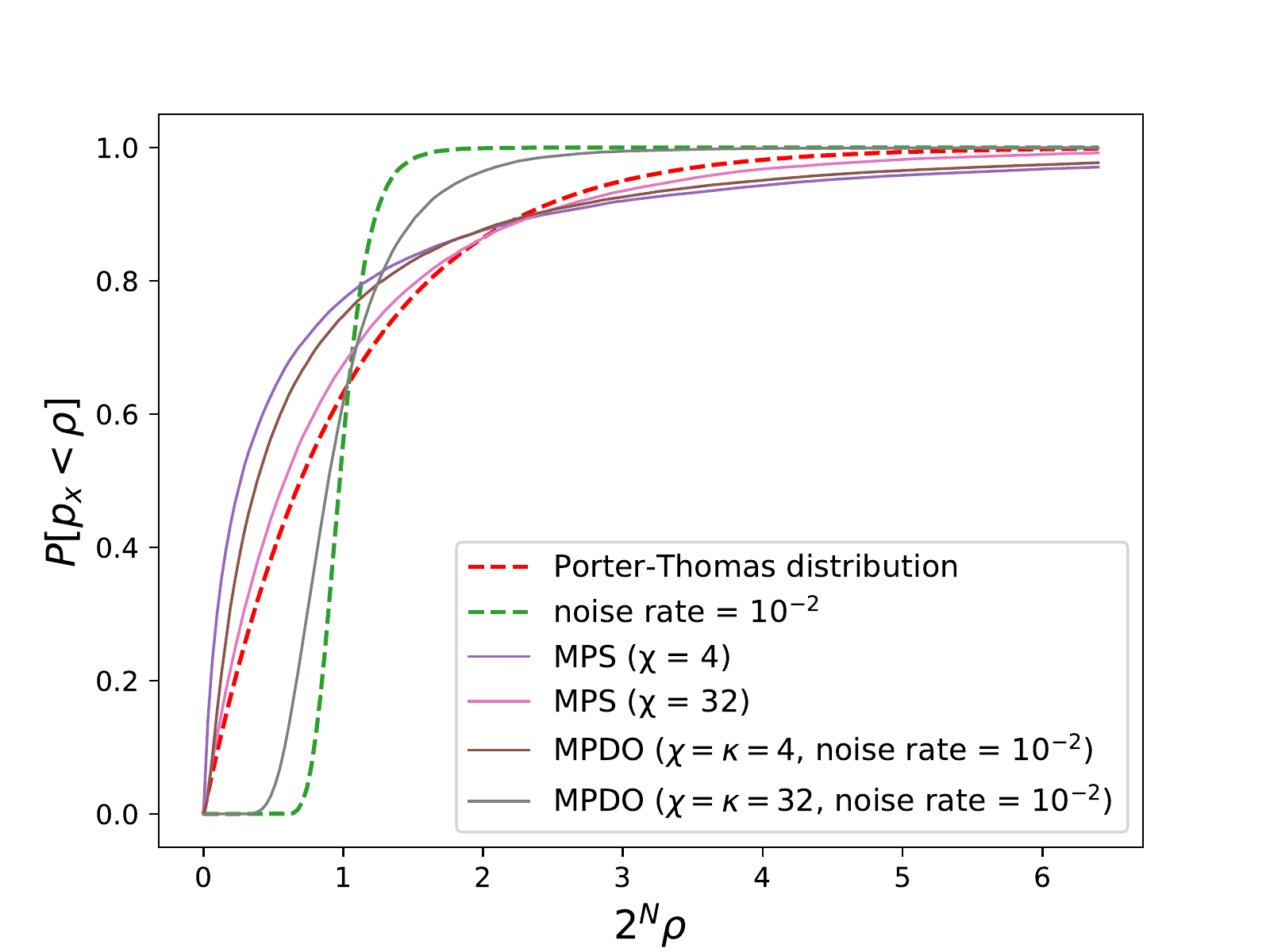}
		\end{minipage}%
	}%
	\subfigure[Depolarizing]{
		\begin{minipage}[t]{0.33\linewidth}
			\centering
			\includegraphics[scale=0.35]{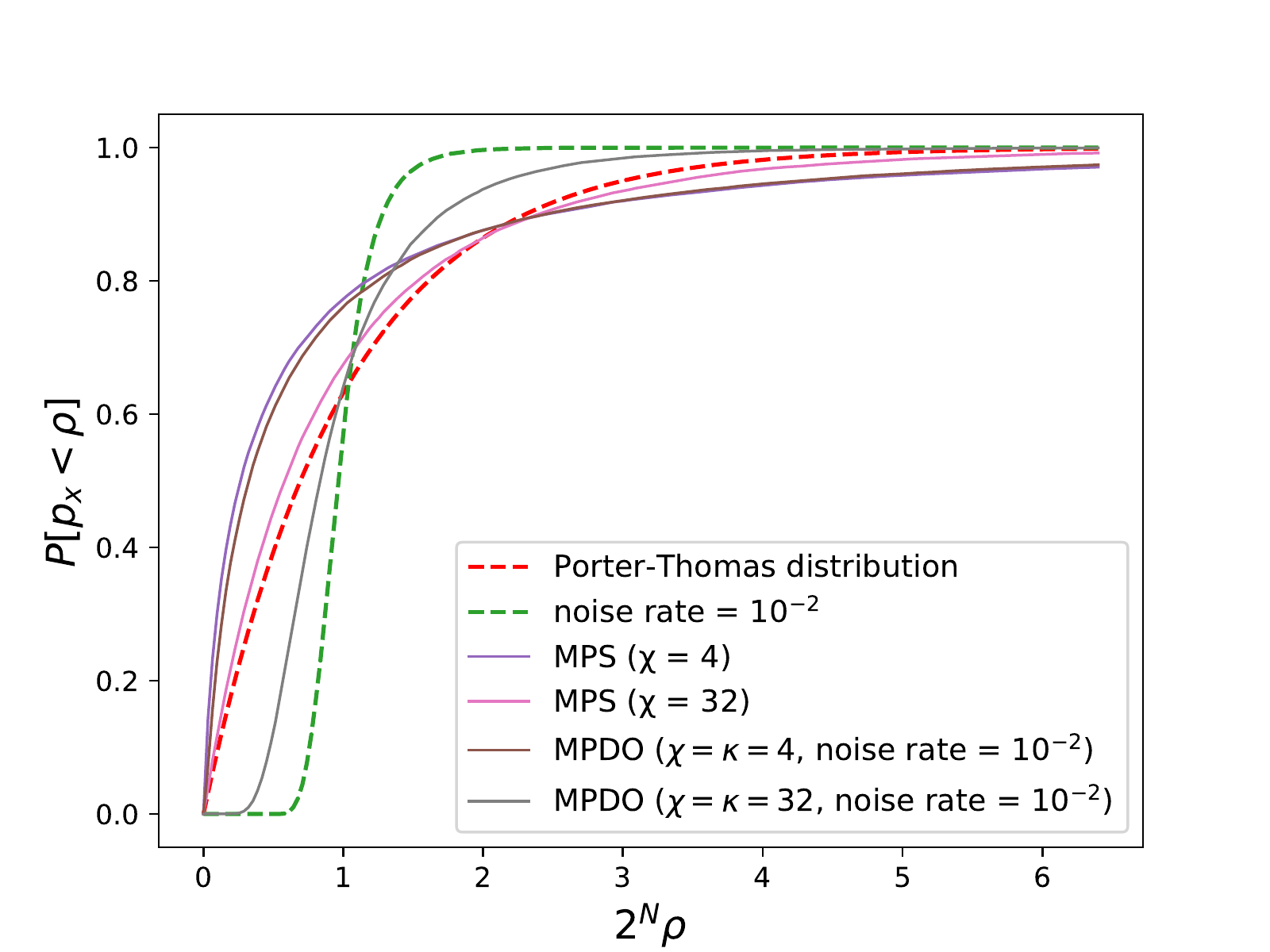}
		\end{minipage}%
	}%
	\subfigure[Amplitude damping]{
		\begin{minipage}[t]{0.33\linewidth}
			\centering
			\includegraphics[scale=0.35]{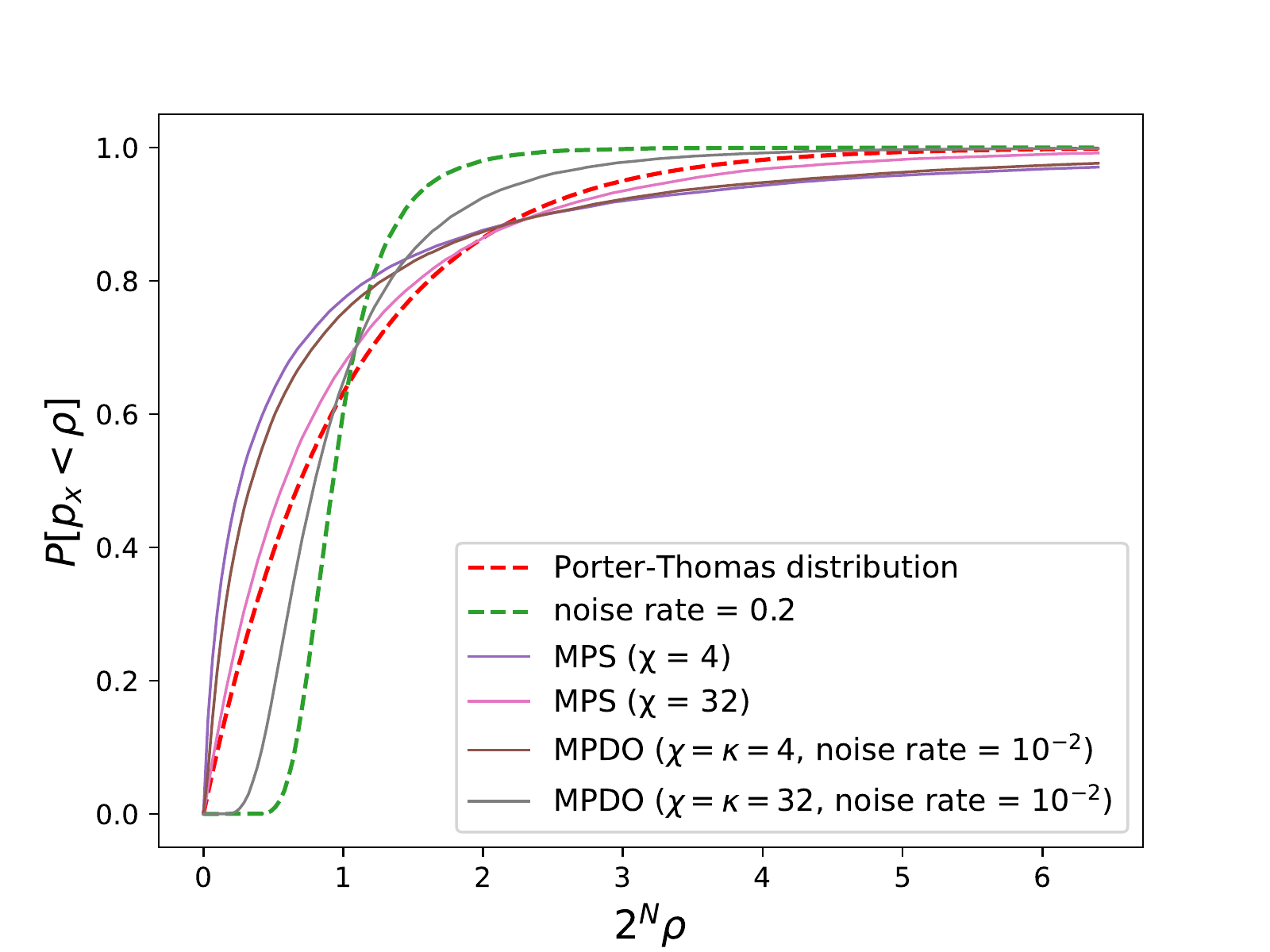}
		\end{minipage}
	}%
	\centering
	\caption{Approximate results of the cumulated $p$ distribution for different noise simulators. The system is 1D random circuit with $15$ qubits and $24$ layers, under three different error models. 
	(a) the dephasing noise; (b) the depolarizing noise; (c) the amplitude damping noise. The green dash line refers to the result of exact simulation. As a signal that MPS cannot faithfully simulated with noise, when $\chi$ increase, the distribution of MPS simulation will only get closer and closer to the Porter-Thomas distribution, while MPDO can gradually approach the correct cumulated $p$  distribution.}
	\label{fig:pt}
\end{figure*}

For a general random circuits, when $D \sim N$, the output states would reach to the Porter-Thomas distribution. However, in real world, certain physical foundation would cause certain type of noises, which leads to the deviation of output states from the Porter-Thomas distribution. This deviation should be correctly captured by a proper noisy simulator.

In this section, we consider random circuits with $15$ qubits and depth $D=24$. We focus on analyzing
how the Porter-Thomas distribution changes due to the effect of noise, and compare the different method
of simulation. 

For a density matrix $\rho$, consider a random variable $p_i=\langle x_i |\rho|x_i\rangle$ for $x_i$ is the $i$-th bit-string from $\{0,1\}^N$, thus $|x_i\rangle$ is one of the computational basis. 

If there is no noise, the output pure state $\ket{\psi}$ is resulted from the random circuit $U$:$\ket{\psi}=U\ket{0}$.
Then for a pure state, the probability of getting a certain base $\ket{x_i}$ is
\begin{equation}
p_i=|c_i|^2=|\bra{x_i}U\ket{0}|^2.
\end{equation}
For a random circuit with sufficiently large depth, the distribution of $\{p=p_i(x)\}$ is known to follow the Porter-Thomas distribution,
\begin{equation}
\label{eq:ptd}
Pr(p)=(M-1)(1-p)^{M-2}=M\left(e^{-Mp}+O(1)\right),
\end{equation}
with expectation $1/M$~\cite{boixo2018characterizing}, where $M=2^N$.

For random with $15$ qubits and depth $24$, we calculate the cumulated $p$ distribution for different noise models with:
\begin{itemize}
\item The Porter-Thomas distribution (corresponding to a exact noiseless simulator)
\item Distribution given by an exact noise simulator.
\item Distribution given by an MPS simulator.
\item Distribution given by an MPDO simulator discussed in Sec.~\ref{sec:MPDO}.
\end{itemize}

In Fig.~\ref{fig:pt_exact}, we compare the Porter-Thomas distribution with exact simulations of different cumulated $p$ distributions under three different type of noise to show the derivation of the output from the Porter-Thomas distribution. As a qualititively comparison, we include some analytical discussion for a simplified depolarizing noise model in Appendix A. 

Results of approximated simulation are summarized in Fig.~\ref{fig:pt}, which clearly shows that the MPDO distribution can approach its corresponding exact noisy distribution with the increase of the bond dimension $\chi$ and the inner dimension $\kappa$. 
While the MPS method does not approximate the actual output distribution of $\rho_e$. Notice that a relatively small bond and inner dimensions $\chi=\kappa=32$ for the MPDO simulation already grasp some qualitative behavior of the cumulated $p$ distribution.  \\

\section{Truncation of bond and inner dimensions}
\label{sec:truncation}

\begin{figure*}[htbp]
	\centering
	\subfigure[Dephasing]{
		\begin{minipage}[t]{0.33\linewidth}
			\centering
			\includegraphics[scale=0.35]{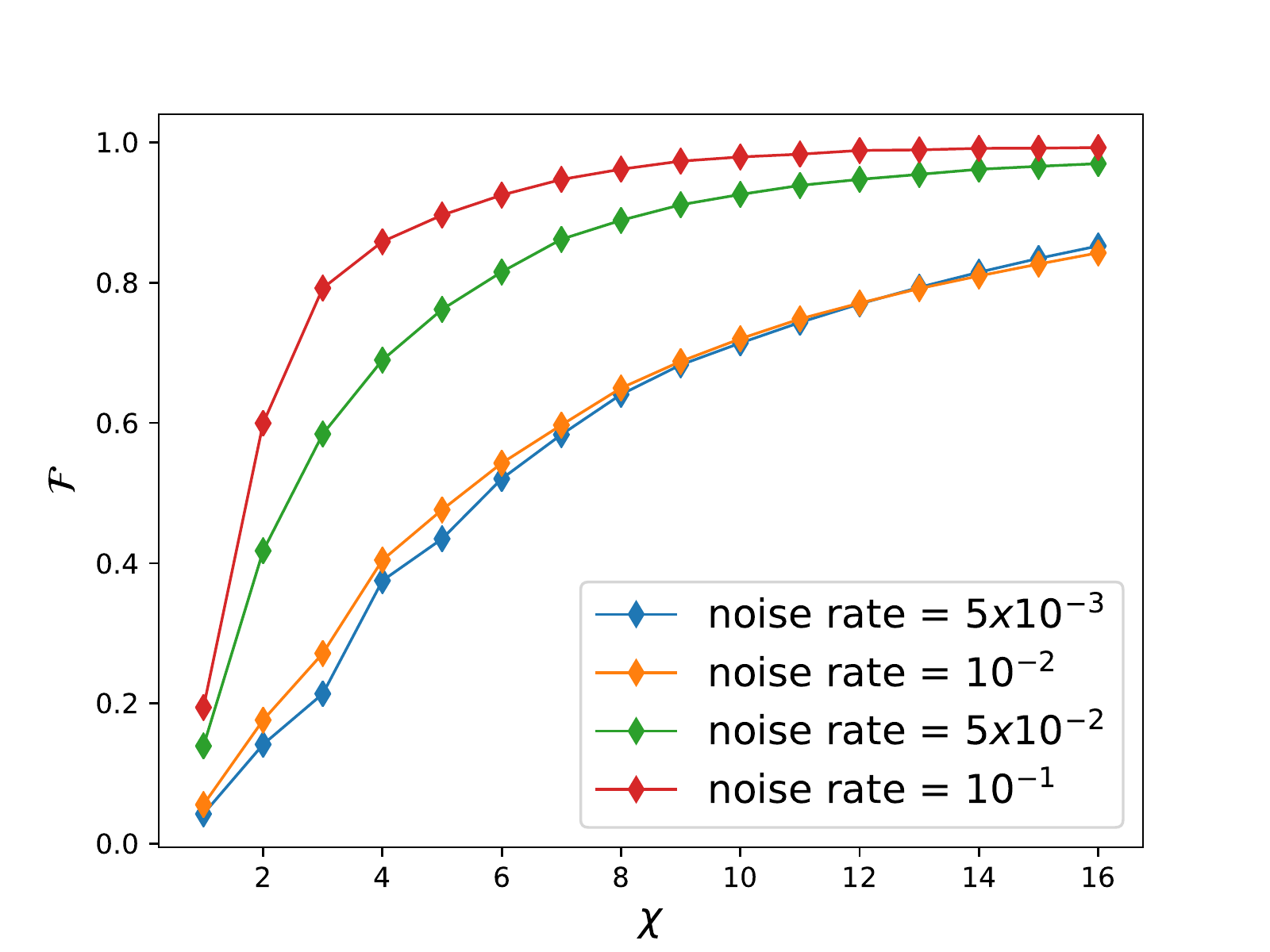}
		\end{minipage}%
	}%
	\subfigure[Depolarizing]{
		\begin{minipage}[t]{0.33\linewidth}
			\centering
			\includegraphics[scale=0.35]{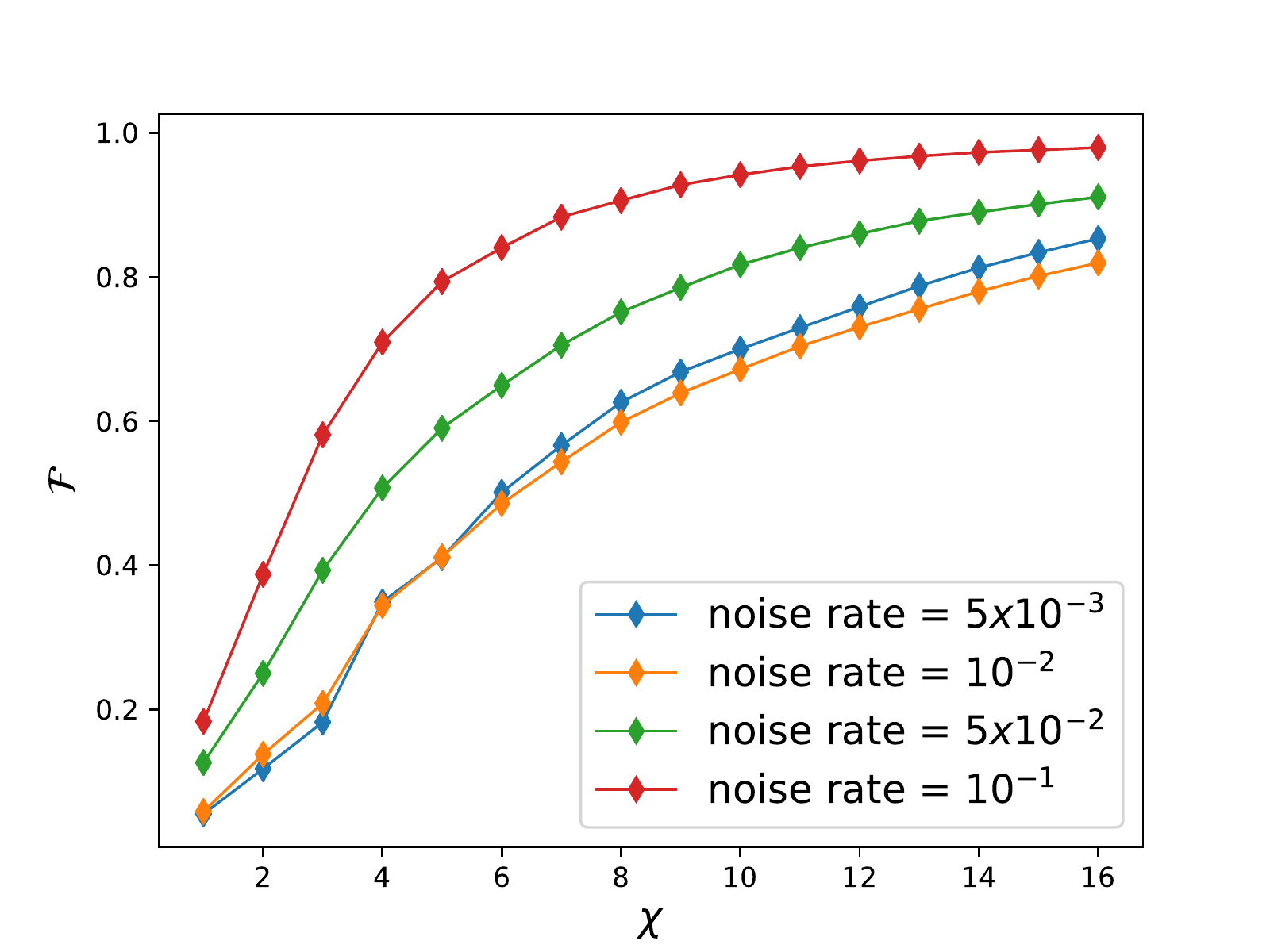}
		\end{minipage}%
	}%
	\subfigure[Amplitude damping]{
		\begin{minipage}[t]{0.33\linewidth}
			\centering
			\includegraphics[scale=0.35]{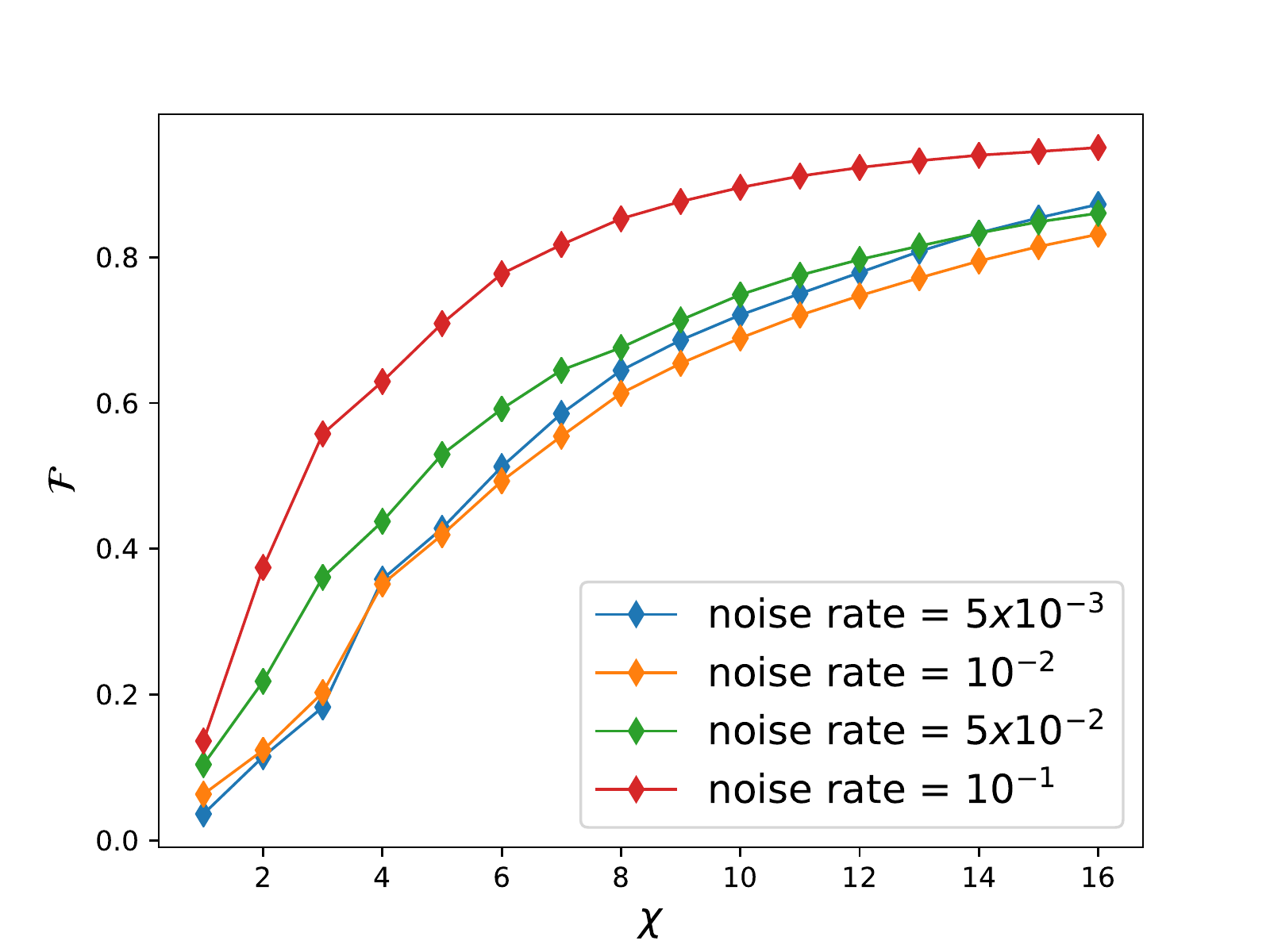}
		\end{minipage}
	}%
	\centering
	\caption{$\mathcal{F}(\rho_{e}, \rho_{d})$ for different noise rates and different MPDO with maximum bond dimension $\chi$ and maximum inner dimension $\kappa=2 \chi$. (a) the dephasing noise; (b) the depolarizing noise; (c) the amplitude damping noise. }
	\label{fig:noise_bond}
\end{figure*}

In this section, we study the effect of truncation in our MPDO for bond and inner dimensions. We consider random circuits with $10$ qubits and depth $D=24$. 

We first study the effect of truncating the bond dimension. Notice bond dimension, in fact, puts an upper bound for the inner dimension. That is, when the bond dimension is truncated to a maximum value of $\chi$, the inner dimension is upper bounded by $2\chi^2$. For each of the error models with different error rates, we choose to truncate the bond dimension to a maximum value $\chi$ and also truncate the inner dimension to a maximum value $\kappa = 2\chi$. and computes $\mathcal{F}(\rho_{e}, \rho_{d})$, to see how well the MPDO method approximates the exact noisy result $\rho_e$.
Our results are shown in Fig.~\ref{fig:noise_bond}.

As shown in Fig.~\ref{fig:noise_bond}, when the gate error is larger than some threshold (approximately $0.01$ for all the error models), smaller bound dimension suffices to simulate the noisy circuits, indicating it is the regime that the noise is large enough for the quantum circuit to be `easy' for classical simulation.

\begin{figure*}[htbp]
	\centering
	\subfigure[Dephasing]{
		\begin{minipage}[t]{0.33\linewidth}
			\centering
			\includegraphics[scale=0.35]{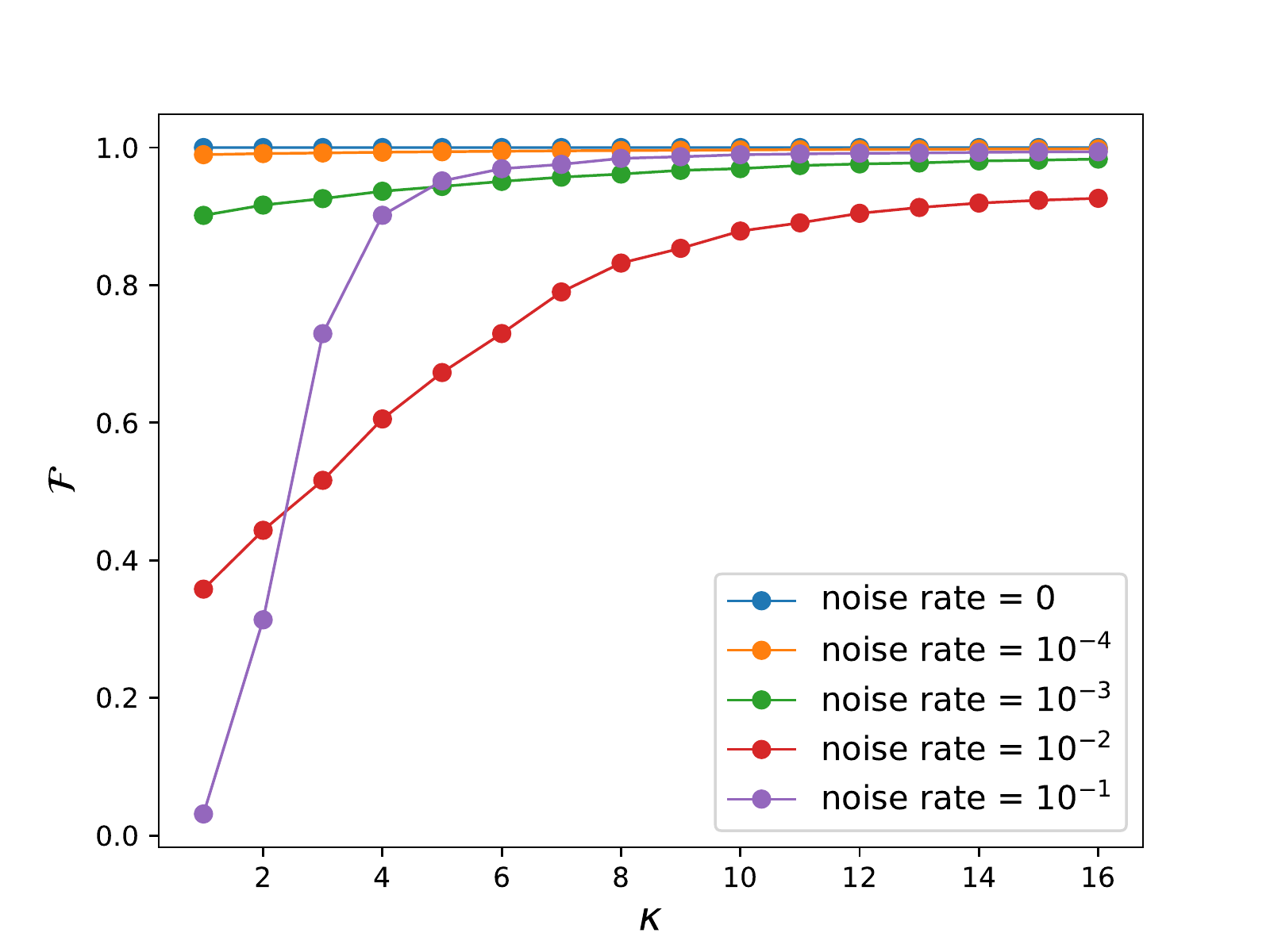}
		\end{minipage}%
	}%
	\subfigure[Depolarizing]{
		\begin{minipage}[t]{0.33\linewidth}
			\centering
			\includegraphics[scale=0.35]{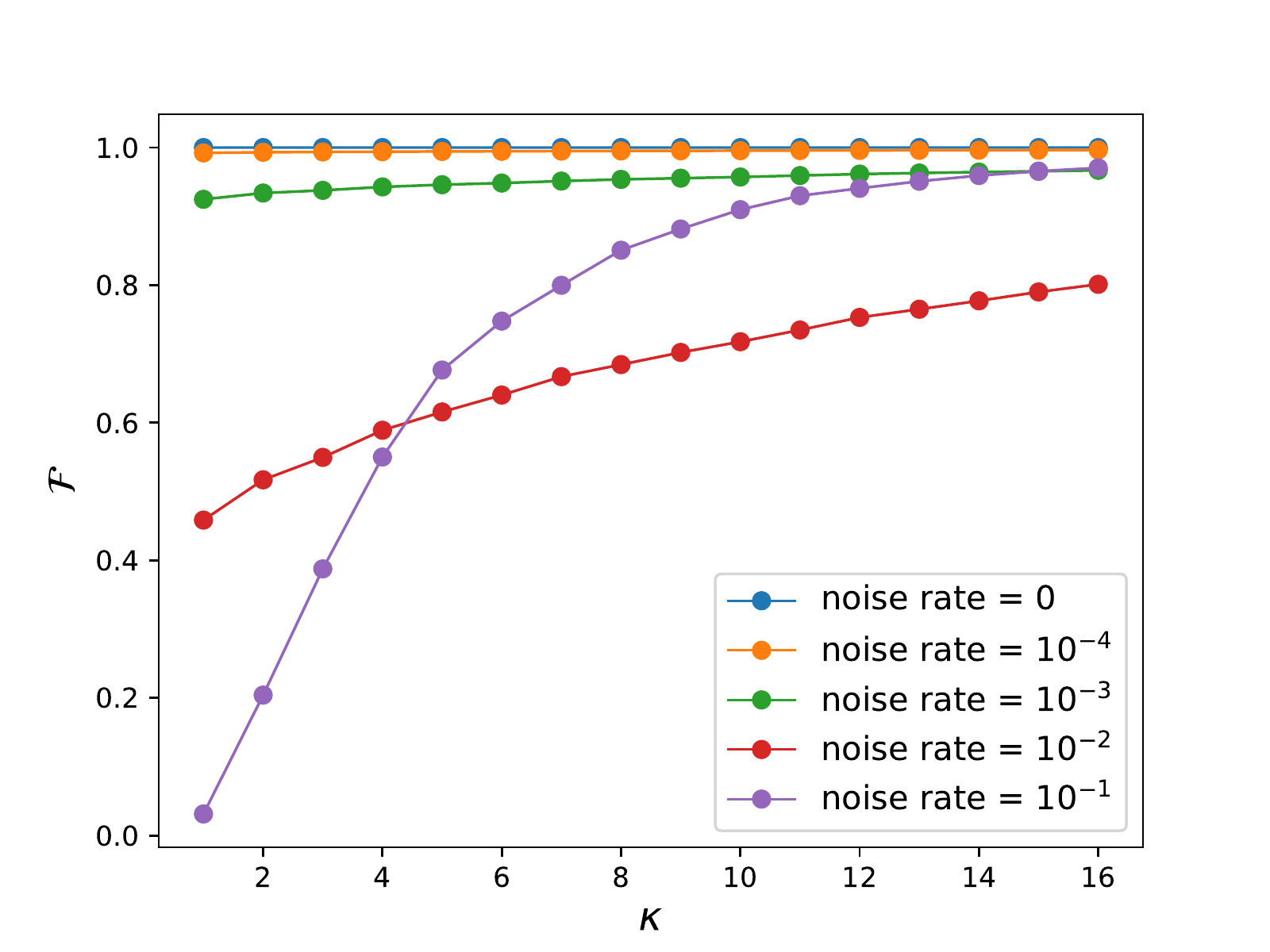}
		\end{minipage}%
	}%
	\subfigure[Amplitude damping]{
		\begin{minipage}[t]{0.33\linewidth}
			\centering
			\includegraphics[scale=0.35]{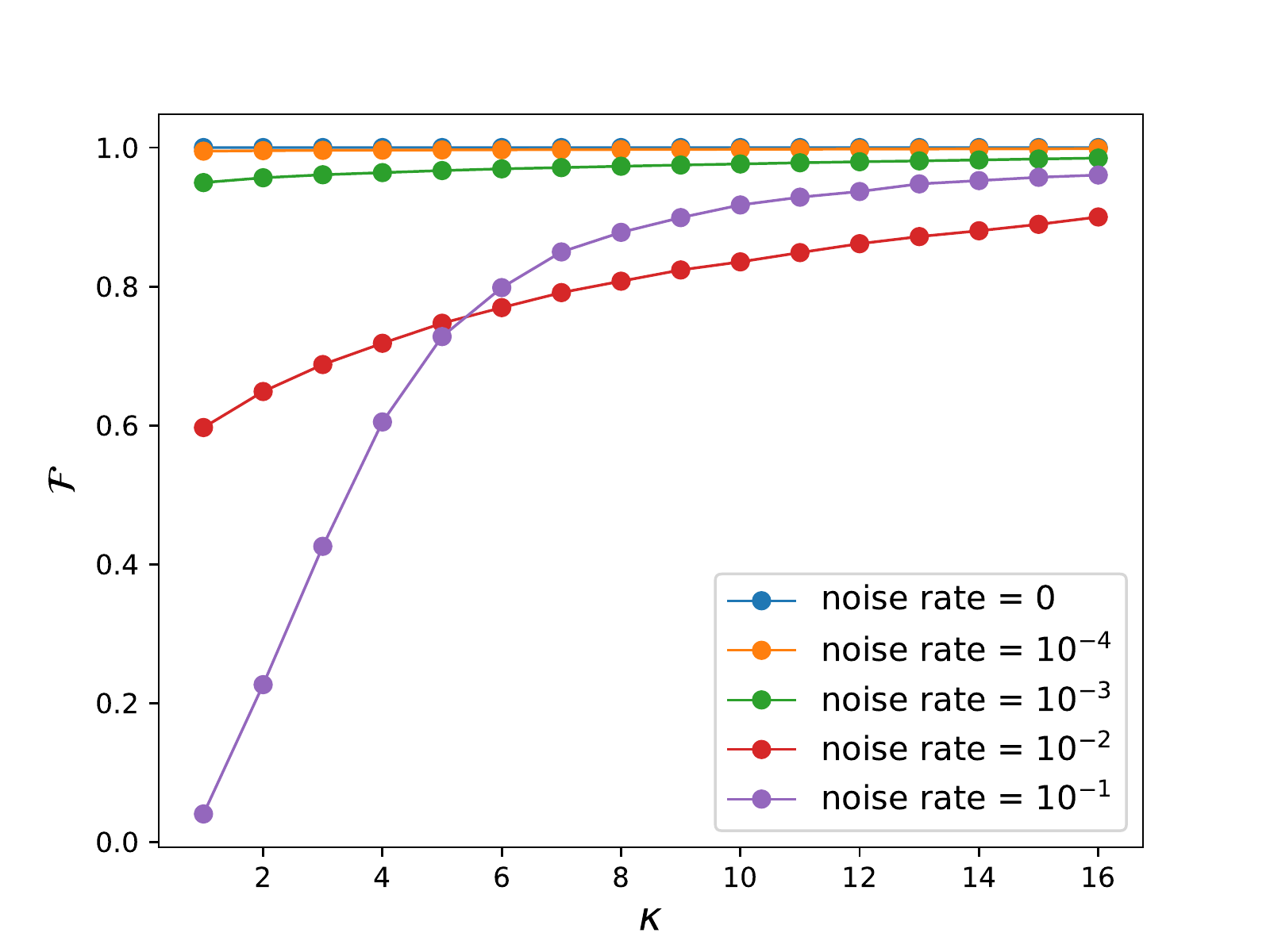}
		\end{minipage}
	}%
	\centering
	\caption{$\mathcal{F}(\rho_{e}, \rho_{d})$ for different noise rates and different MPDO with inner dimension $\kappa$. (a) the dephasing noise; (b) the depolarizing noise; (c) the amplitude damping noise. }
	\label{fig:inner_bond}
\end{figure*}

We further study the effect of truncating the inner dimensions. For each of the error models with different error rates, we choose to truncate the bond dimension to a maximum value $\chi=32$ and to truncate the inner dimension to a maximum value of $\kappa$. We compute $\mathcal{F}(\rho_{e}, \rho_{d})$, to see how well the MPDO method approximates the exact noisy result $\rho_e$.
Our results are shown in Fig.~\ref{fig:inner_bond}. 

As shown in Fig.~\ref{fig:inner_bond}, in case of weak system noise, smaller inner dimension suffices to simulate the noisy circuits. In this case, the memory cost of MPDO is significantly less than the MPO method by directly using the $M$ tensor.

\section{Experiments on IBM quantum devices}
\label{sec:IBM}
To test our MPDO method with real quantum computers, we 
run several 1D random circuits on the IBM device. We use the $15$-qubit device ibmq$\_$16$\_$melbourne~\cite{IBMQ2020}. The structure of ibmq$\_$16$\_$melbourne is shown in FIG. 8(b). 

We run $10$-qubit random circuits on a chain, which consists of the qubits $0, 1, 2, 3, 4, 5, 6, 8, 9, 10$, and simulate these circuits with our MPDO method. The parameters of ibmq$\_$16$\_$melbourne are given in Table~\ref{tab:my_label}. The rightmost column shows the CNOT error rates. cxi\_j represents the error rate for CNOT operation of control qubit-i and target qubit-j. cxi\_j = cxj\_i always holds, so we just list one of them.

\begin{figure}[hbtp]
\centering
    \includegraphics[scale=0.3]{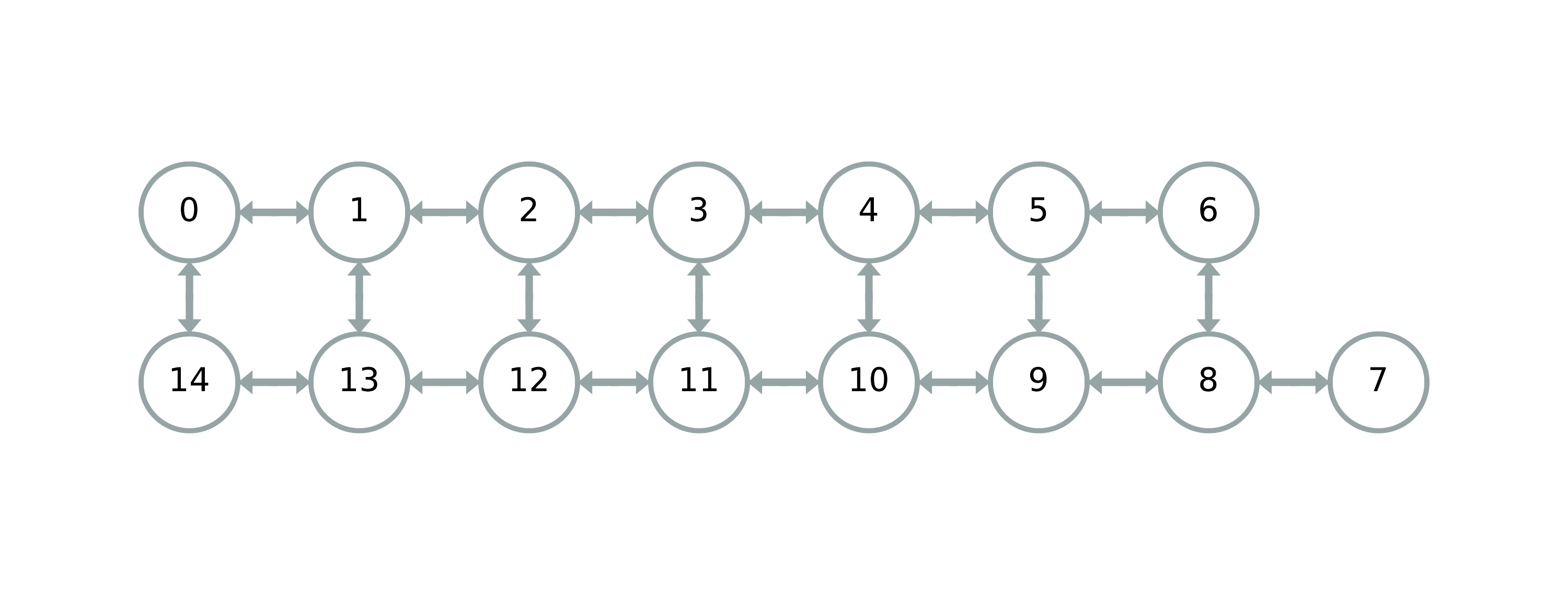}
    \caption{Structure of ibmq$\_$16$\_$melbourne}
    \label{fig:ibmq}
\end{figure}

\begin{figure*}[htbp]
	\centering
	\subfigure[~Cross entropy $H$ between experimental results and simulation results of MPS with maximum bond dimension $\chi$ for circuits with $D$ layers.]{
		\begin{minipage}[t]{0.49\linewidth}
			\centering
			\includegraphics[scale=0.45]{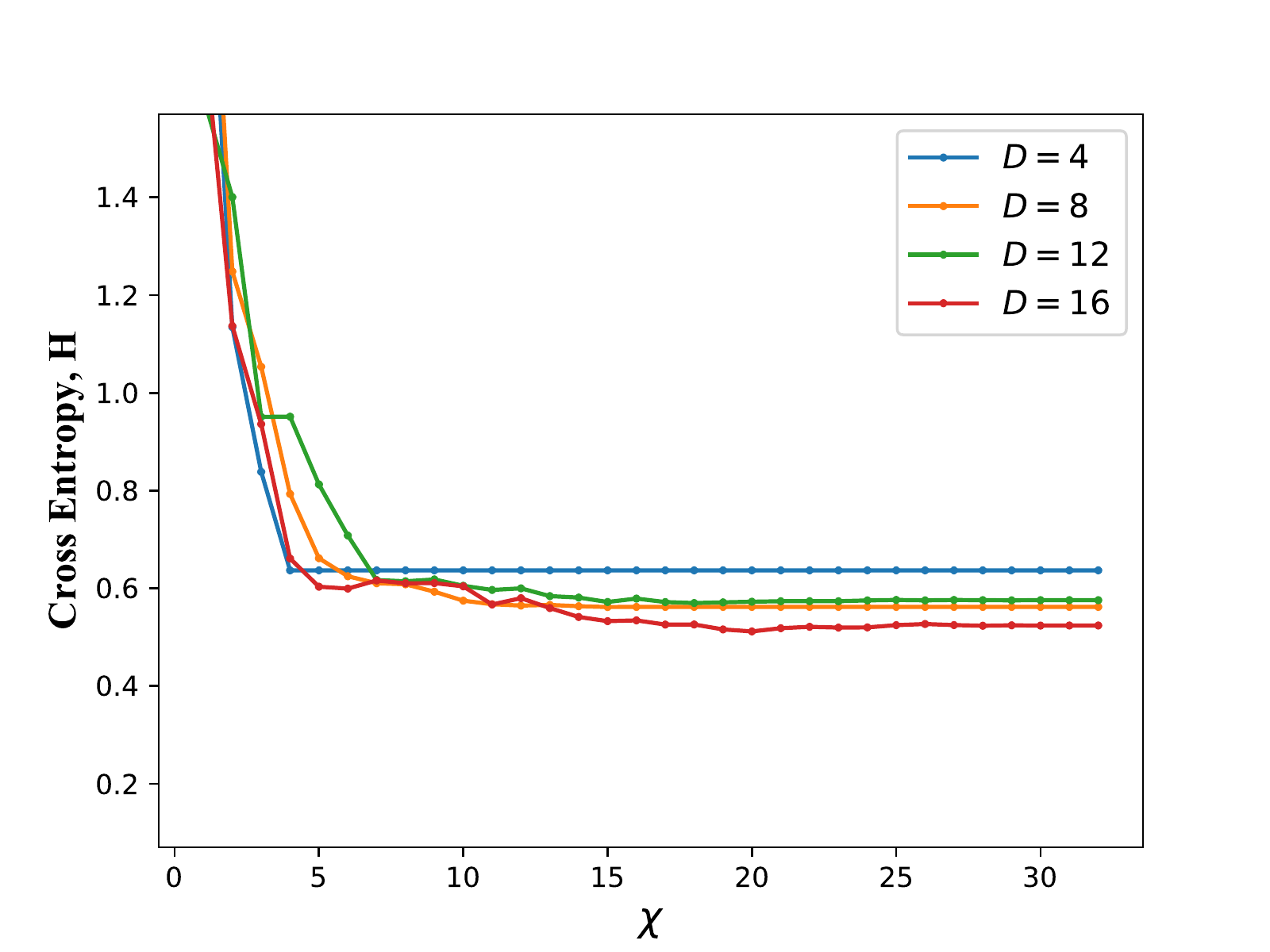}
		\end{minipage}%
	}%
	\subfigure[~Cross entropy $H$ between experimental results and simulation results of MPDO with maximum bond dimension $\chi$  and maximum inner dimension $\kappa = 2 \chi$ for circuits with $D$ layers. \textbf{Inset}: Cross entropy $H$ between experimental results and simulation results of MPDO with maximum inner dimension $\kappa$ and maximum bond dimension $\chi = 32$ for circuits with $D$ layers.]{
		\begin{minipage}[t]{0.49\linewidth}
			\centering
			\includegraphics[scale=0.45]{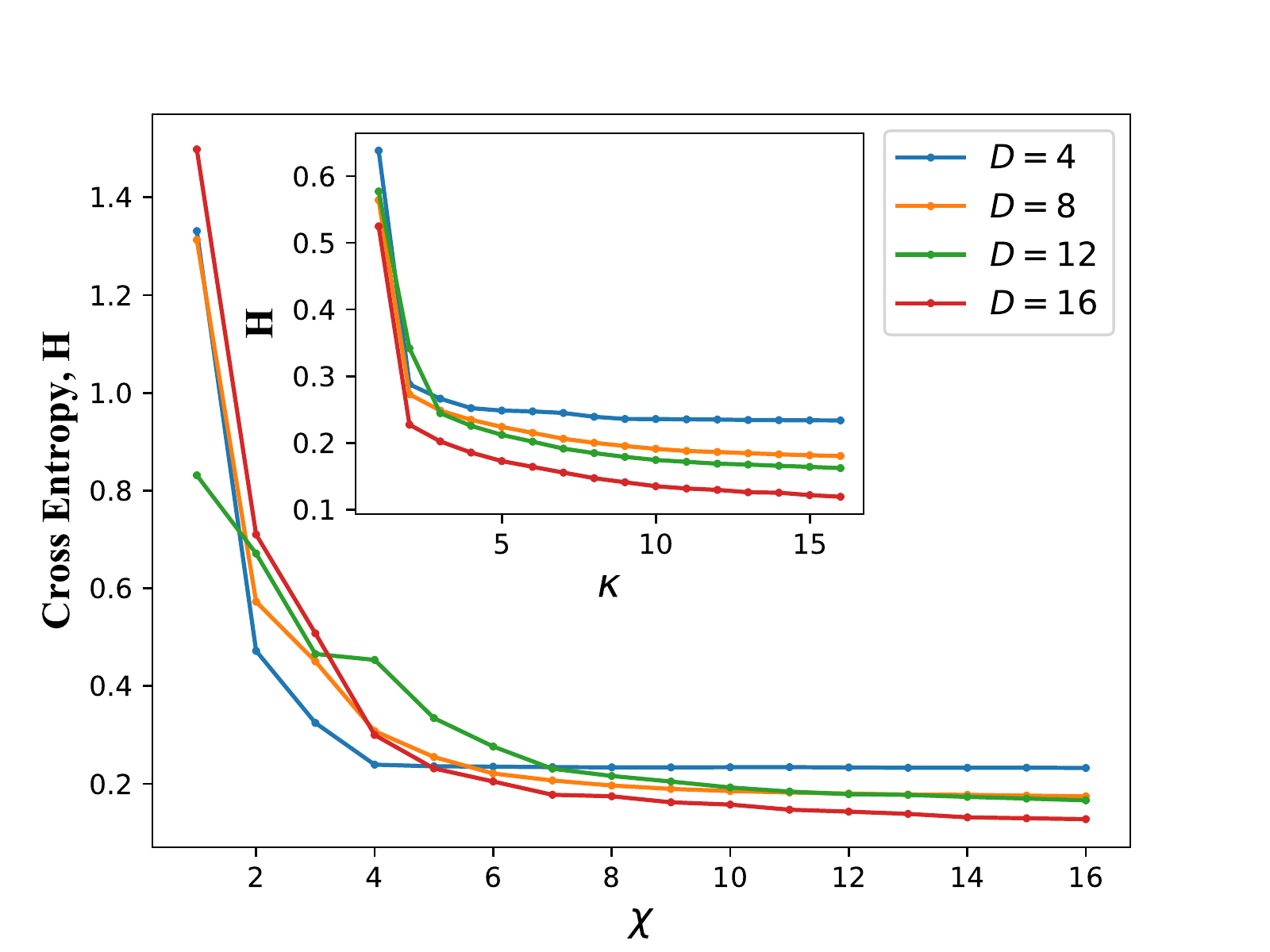}
		\end{minipage}
	}%
	\centering
	\caption{Cross entropy between simulations and experiments on the ibmq$\_$16$\_$melbourne device.}
	\label{fig:cross_entropy}
\end{figure*}

\begin{table}[htbp]
	\centering
	\begin{tabular}{ |p{1cm}||p{1.3cm}|p{2cm}|p{3cm}|   }
		\hline
		Qubit & Readout error  & Single-qubit U2 error rate & CNOT error rate \\
		\hline
		0   & 0.0185  & $5.48 \times 10^{-4} $ & cx0\_1: 0.0236 \\
		1   & 0.0915 & $2.78 \times 10^{-3} $ & cx1\_2: 0.0165 \\
		2   & 0.0395 & $8.90 \times 10^{-4} $& cx2\_3: 0.0171 \\
		3  & 0.0475 & $3.78 \times 10^{-4}$ & cx3\_4: 0.0169 \\
		4   & 0.0595 & $1.06 \times 10^{-3}$ &cx4\_5: 0.0295\\
		5  & 0.0615 & $2.34 \times 10^{-3}$ &cx5\_6: 0.0467\\
		6   & 0.027 &$1.27 \times 10^{-3} $&cx6\_8: 0.0322\\
		8   & 0.283 & $8.13 \times 10^{-4} $&cx8\_9: 0.0346\\
		9   & 0.05 & $9.66 \times 10^{-3} $&cx9\_10: 0.0510\\
		10   & 0.03 & $1.60 \times 10^{-3} $&\\
		
		\hline
	\end{tabular}\\
	\caption{Noise rates of ibmq$\_$16$\_$melbourne}
	\label{tab:my_label}
\end{table}

Considering three $10$-qubit random circuits with $D=4, 8, 12, 16$ layers respectively. Each circuit is run and measured in the computational basis on ibmq$\_$16$\_$melbourne for $8192$ times. We assume the noise model is depolarizing noise, and simulate these circuits via MPDO according to the noise rates given in Table~\ref{tab:my_label}. 

Denote $P({x}_{i})$ as the probability of bitstring $x_i$ in experiment, and $P_{s}({x}_{i})$ as the probability of $x_i$ in our MPDO simulation. To measure the similarity between $P$ and $P_{s}$, we use the cross entropy between the distributions $P$ and $P_{s}$ as given by
\begin{equation}
H(P, P_{s})=-\sum P(x) \log P_{s}(x)
\end{equation}
We first show the results between MPS models and experiments in Fig.~\ref{fig:cross_entropy}(a). we truncate the bond dimension of MPS to a maximum value $\chi$. However, the cross entropy of the MPS model no longer drop with increasing $\chi$ when it reaches $\sim0.5$, which indicated that the MPS model fails to express the real physical noise in quantum circuits.
For our MPDO simulation, we truncate the bond dimension to a maximum value $\chi$ and also truncate the inner dimension to a maximum value $\kappa = 2\chi$. The results of $H(P, P_{s})$ are shown in Fig.~\ref{fig:cross_entropy}(b). Then we truncate the bond dimension $\chi$ to a maximum value $32$ and the inner dimension to a maximum value $\kappa$, the results of cross entropy $H(P, P_{s})$ versus $\kappa$ are shown in the inset of Fig.~\ref{fig:cross_entropy}(b). These results not only show that the MPDO model is better than the MPS model in simulating a real quantum computer, but also shows that relatively small $\chi$ and $\kappa$ can already simulate the noisy random circuit efficiently.

\section{Simulating Encoding Circuits for Quantum Error-Correcting Codes}
\label{sec:QEC}

\begin{figure*}[htbp]
	\centering
	\subfigure[Dephasing]{
		\begin{minipage}[t]{0.49\linewidth}
			\centering
			\includegraphics[scale=0.45]{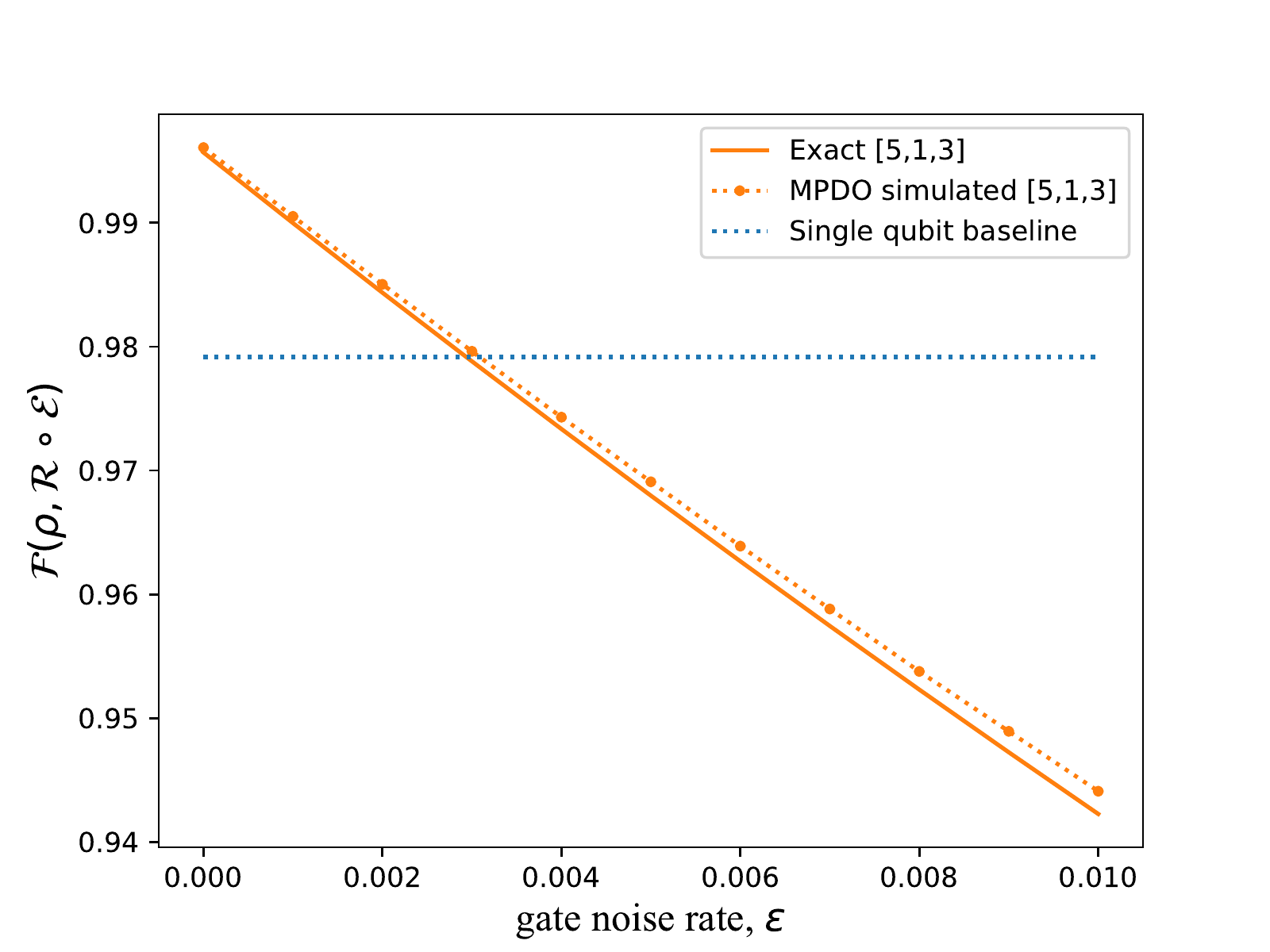}
		\end{minipage}%
	}%
	\subfigure[Depolarizing]{
		\begin{minipage}[t]{0.49\linewidth}
			\centering
			\includegraphics[scale=0.45]{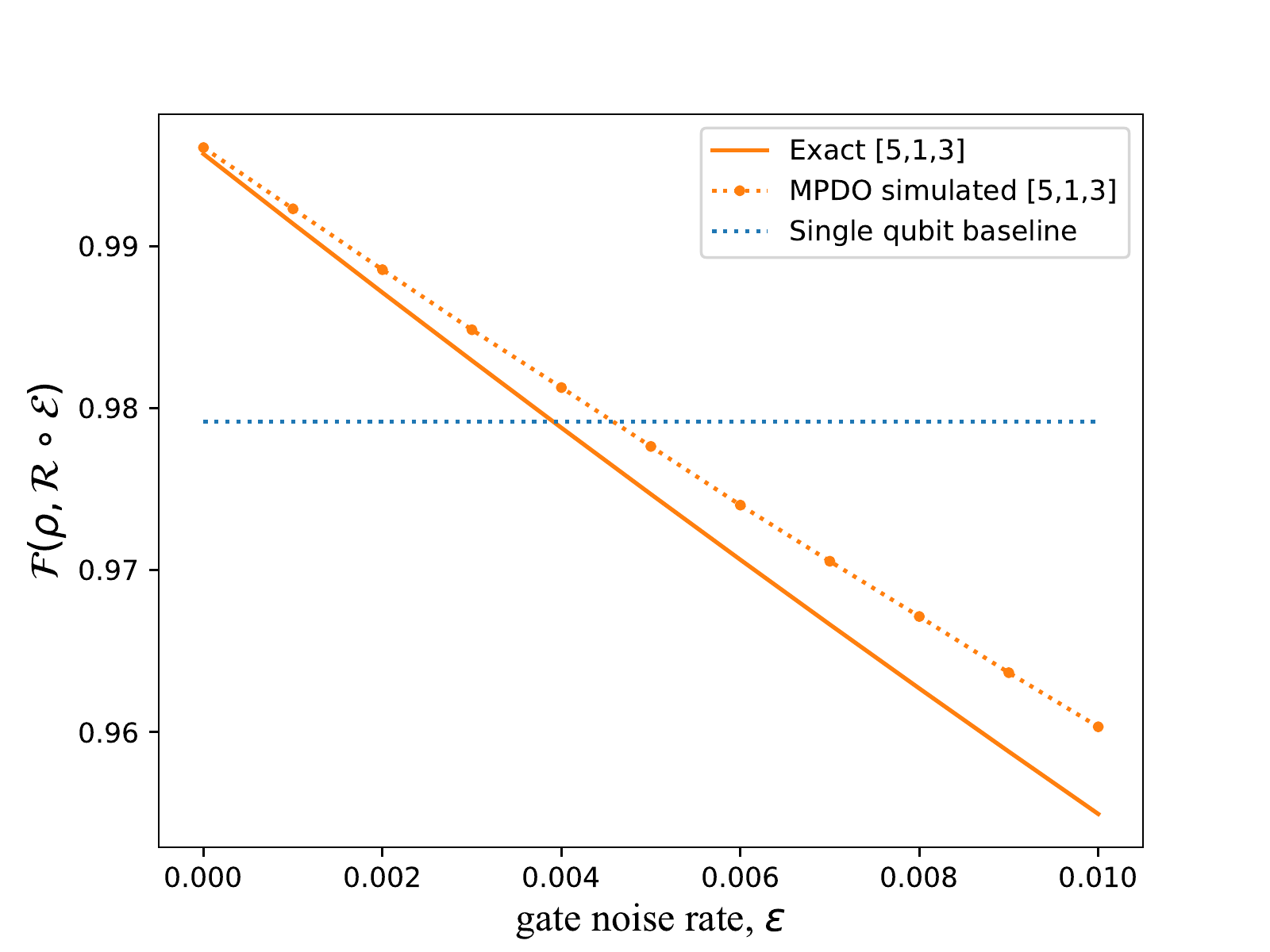}
		\end{minipage}%
	}%
	\vskip\baselineskip
	\subfigure[Amplitude damping]{
		\begin{minipage}[t]{0.49\linewidth}
			\centering
			\includegraphics[scale=0.45]{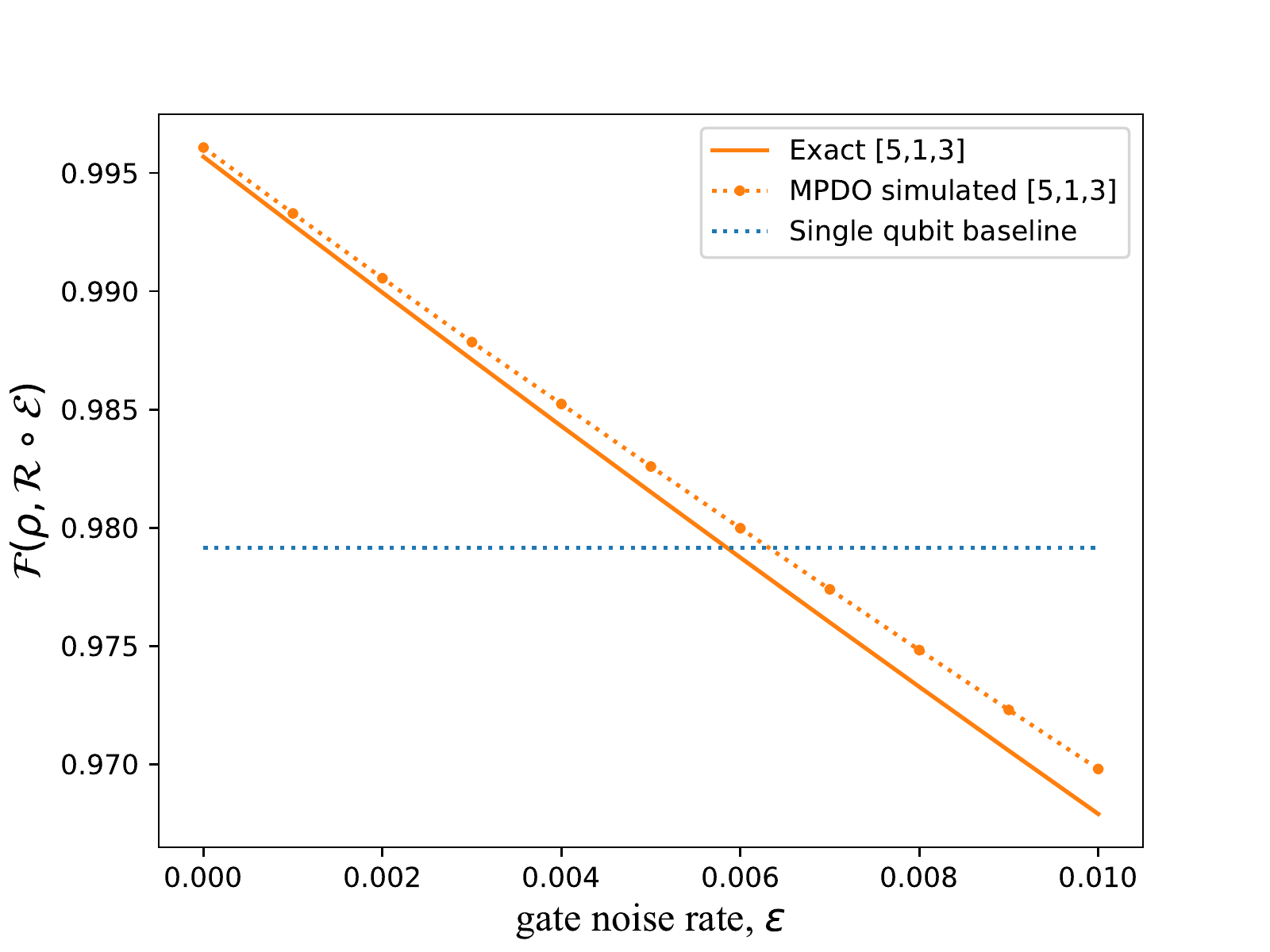}
		\end{minipage}
	}%
	\subfigure[Colelctive dephasing]{
	\begin{minipage}[t]{0.49\linewidth}
		\centering
		\includegraphics[scale=0.45]{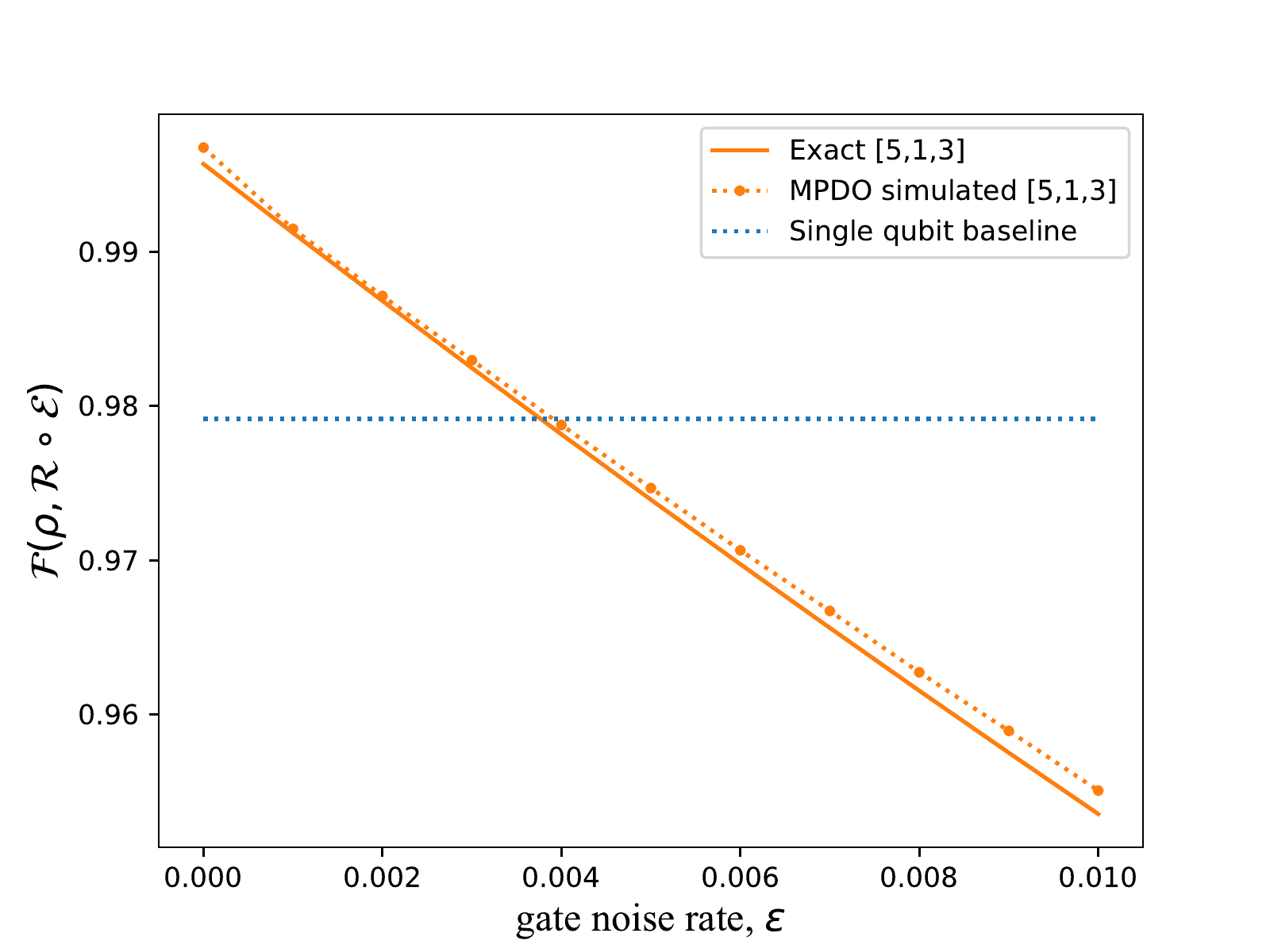}
	\end{minipage}
}%
	\centering
	\caption{Recovered fidelity versus gate noise $\epsilon$ for [5,1,3] perfect code. The blue line is the fidelity without quantum error correction. (a) the dephasing gate noise; (b) the depolarizing gate noise; (c) the amplitude damping gate noise. (c) the collective dephasing gate noise.}
	\label{fig:qec}
\end{figure*}

\begin{figure}[hbtp]
	\centering
	\includegraphics[scale=0.28]{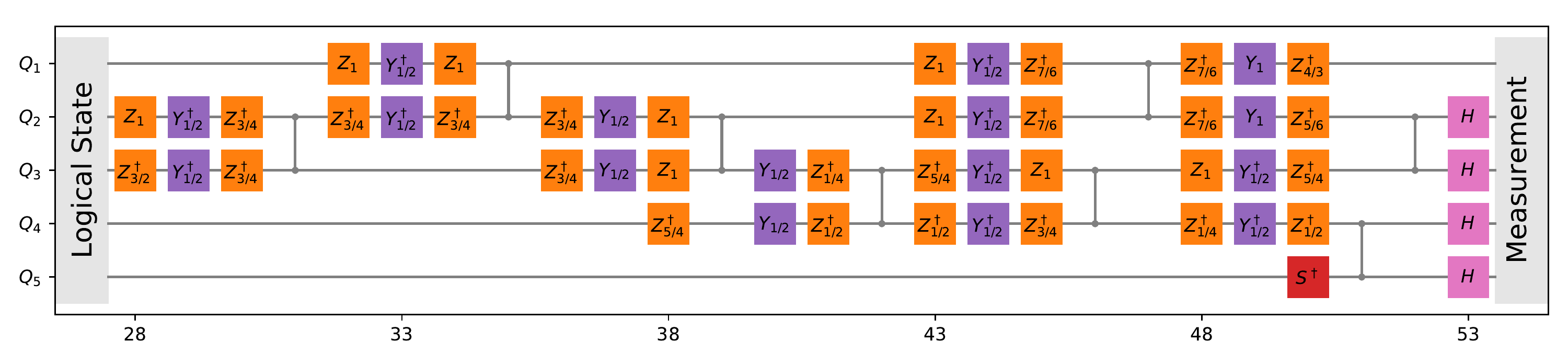}
	\caption{Encoder of the [5,1,3] code}
	\label{fig:encoder}
\end{figure}

Apart from random circuits, the tensor-networks nature of the MPDO method guarantee a general good performance when applying to various systems with small entanglement or local correlations, which in fact include the most common circuits in the NISQ era~\cite{Verstraete:2008ko,2011AnPhyS,Orus2014,Pirvu_2010}. Besides, just as the two-dimensional DMRG algorithm~\cite{2011AnPhyS}, this method could also be directly applied to the two-dimensional system.

In this section, we give an example of applying the MPDO method to simulate encoding circuits for quantum error-correction codes. We consider the 5-qubit code~\cite{2019arXiv190704507G}, whose encoding circuit is shown in Fig.~\ref{fig:encoder}.

For a  single-qubit state in the ensemble 
$\{|0\rangle, |1\rangle, (|0\rangle+|1\rangle)/\sqrt{2}, (|0\rangle-|1\rangle)/\sqrt{2}, (|0\rangle+i|1\rangle)/\sqrt{2}, (|0\rangle-i|1\rangle)/\sqrt{2}\}$,
we encode it to 5 qubits via the encoding circuit, apply a local depolarizing noise channel to all 5 qubits with noise rate 0.05, then decode and recover the initial state. Denote the average fidelity between the initial state and the recovered state as $\mathcal{F}(\rho, \mathcal{R}\circ\mathcal{E})$. 

Suppose the CZ gate in the encoder and decoder is noisy with noise rate $\epsilon$, we considered 4 different noise models in total. In addition to the previous dephasing, depolarizing and amplitude damping noise, we also added an example of neighboring two-qubit noise, the collective dephasing noise~\cite{2003LNP62283L}, which is defined as

\begin{equation}
\rho\rightarrow \mathcal{E}_{CD}(\rho)=(1-\epsilon) \rho + \epsilon Z_{c}\rho Z^{\dag}_{c},
\end{equation}
where  $\epsilon \in [0, 1]$, $Z_{c} = \mathrm{diag}{(1, -1, -1, 1)}$.

Applying a two-qubit noise on the MPDO is similar to the single qubit case. We only need to contract those two neighboring tensors $T^{[i]}$ and $T^{[i+1]}$ to a merged tensor $W$ with a physical dimension of 4, and then directly apply the two-qubit noise gate to $W$ as the way of the single-qubit noise gate. Finally, the SVD decomposition is used to separate the $W$ into two new tensors. The product of the inner dimension of those two tensors is equal to $m$ times of the inner dimension of $W$, where $m$ is the number of terms of the noise model.

We simulate the aforementioned quantum error correction process with exact density matrix simulation and MPDO simulation. The maximum bond dim and inner dim in MPDO are $\chi = 16$ and $\kappa = 32$. The results of recovered fidelity versus gate noise $\epsilon$ for different noise models are shown in Fig.~\ref{fig:qec}. The simulation results of the MPDO are close to the exact ones even with a relative small $\chi$ and $\kappa$.

\section{Discussion}
\label{sec:discussion}
In this work, we have developed a method to simulate noisy quantum circuits based on MPDO. We show that our method approximates the noisy output states well. While the method based on MPS bond dimension truncation, failed to approximate the noisy output states for any of the noise model considered, indicating that the MPS method might not represent any local noise model in real physical systems. 

Our MPDO method exhibits the following advantages.
\begin{itemize}
\item It reflects a clear physical picture, with inner indices taking care of noise simulation, and bond indices taking care of two-qubit gate simulation.
\item Both bond and inner dimensions can be truncated using the SVD method, adaptive to the need of different situations of the noise simulation.
\item In case of strong system noise, small bond dimensions are sufficient to simulate the noisy circuits.
\item In case of weak system noise, the memory cost of MPDO is significantly less than the MPO method.
\item With an effective tensor update scheme that truncates the inner dimension up to a maximum value $\kappa$ and bond dimension up to a maximum value $\chi$, performed after each layer of the circuit, the cost of our simulation scales as $\sim ND\kappa^3\chi^3$, for simulating an $N$-qubit circuit with depth $D$.
\item Experimental results on IBM devices demonstrate that relatively small $\chi$ and $\kappa$ can simulate the noisy random circuit efficiently. 
\end{itemize}

It remains an interesting open question to understand further the relationship between bond/inner dimensions and entanglement/classical entropy. It is also highly desired to generalize our method to simulate noisy circuits in two spatial dimensions so that we can more directly compare it to 2D experimental data from e.g. Google's experiments~\cite{arute2019quantum}.

\section*{Acknowledgement} 
\label{sec:acknowledgement}
We acknowledge the use of IBM Quantum services for this work. Song Cheng is supported by the National Science Foundation of China (No. 12004205). Yongxiang Liu is supported by the National Science Foundation of China (No. 11701536). 

\appendix

\section{Analytical study of the deviation caused by the depolarizing noise}
\label{sec:noise_model}

\begin{figure}[hbtp]
\centering
    \includegraphics[scale=0.45]{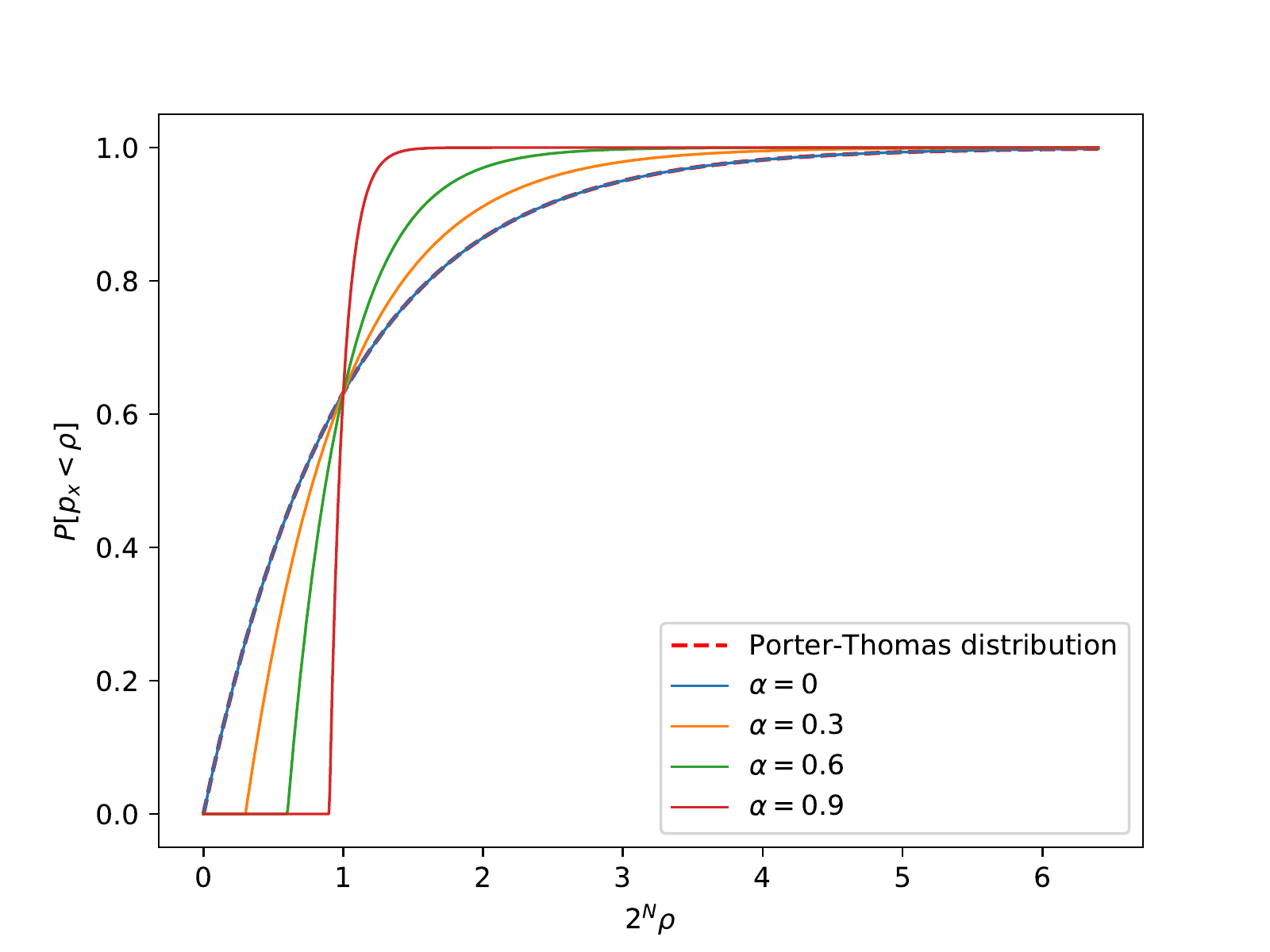}
    \caption{Analysis result of the deviation from Porter-Thomas distribution under the depolarizing noise}
    \label{fig:dp}
\end{figure}

For a simplified model of depolarizing noise, we can also give the analytical form of the deviation from the Porter-Thomas distribution. Consider a simple mode of the depolarizing noise
for a system of $N$ qubits. If the final error rate is $\alpha$, then the final noisy state can be written as
\begin{equation}\label{eq:depolarizing}
\rho=(1-\alpha)\ket{\psi}\bra{\psi}+\alpha\frac{I}{M},
\end{equation}
where $M=2^N$ and $\ket{\psi}$ is the ideal final state with the distribution
\begin{equation}\label{eq:ptd}
Pr(p)=(M-1)(1-p)^{M-2}.
\end{equation}

We notice \eqref{eq:depolarizing}, where the first term $\ket{\psi}\bra{\psi}$ corresponds to the exact output, while the second term 
$\frac{I}{M}$ corresponds to the noise with uniform distribution, which is the white noise.

For the first term, we already known that the random variable $p_1 = |\langle\psi|\phi\rangle|^2$, 
with the state $\ket{\psi}$ chosen 
uniformly at random in the full space, satisfies Porter-Thomas distribution, i.e. \eqref{eq:ptd}. 
For the second term, by similar manipulation, we have that $p_2$ satisfies the single point distribution, 
with the density function to be 
\begin{equation}\label{eq:spd}
Psp(p)=\delta\left(p-\frac{1}{M}\right).
\end{equation}

Then we calculate the probability with noise. Note that $p\in[0,1]$ in \eqref{eq:ptd} and \eqref{eq:spd}, however, 
for convenience, we should make some extension of the function $Pr(p)$ and $Psp(p)$. We define 
\begin{equation*}
\overline{Pr}(p):=\left\{
\begin{split}
&0,  &\quad \text{if }  p<0,\\
&(M-1)(1-p)^{M-2}, &\quad\text{if } 0\leqslant p\leqslant 1, \\
&0,  &\quad\text{if } p>1.
\end{split}
\right.
\end{equation*}
For $Psp(p)$, we directly make the 0 extension.  

According to \eqref{eq:depolarizing}, since the assumption that white noise is independent of the exact result,
 the probability with noise should be
 \newline
\begin{widetext} 
\begin{equation}
\begin{split}
&P((1-\alpha) p_1+\alpha p_2<p)\\
&=\iint\limits_{(1-\alpha) x+\alpha y<p}\overline{Pr}(x)\overline{Psp}(y)dxdy\\
&=\int_{-\infty}^{+\infty}\int_{-\infty}^{\frac{p-\alpha y}{1-\alpha}} \overline{Pr}(x)\delta\left(y-\frac{1}{M}\right)dxdy\\
&=\int_{-\infty}^{+\infty}\int_{-\infty}^{p}\frac{1}{1-\alpha}\overline{Pr}\left(\frac{z-\alpha y}{1-\alpha}\right)\delta\left(y-\frac{1}{M}\right)dzdy\\
&=\int_{-\infty}^{p}\int_{-\infty}^{+\infty}\frac{1}{1-\alpha}\overline{Pr}\left(\frac{z-\alpha y}{1-\alpha}\right)\delta\left(y-\frac{1}{M}\right)dydz\\
&=\int_{-\infty}^{p}\frac{1}{1-\alpha}\overline{Pr}\left(\frac{z-\alpha/M}{1-\alpha}\right)dz
\end{split}
\end{equation}
 
By further calculation, we obtain the result
\begin{equation}
P((1-\alpha) p_1+\alpha p_2<p)=\left\{
\begin{split}
&0,  &\quad \text{if }  0\leqslant p\leqslant \frac{\alpha}{M},\\
&1-\left(1-\frac{p-\alpha/M}{1-\alpha} \right)^{M-1}, &\quad\text{if } \frac{\alpha}{M}< p
< (1-\alpha)+\frac{\alpha}{M}, \\
&1,  &\quad\text{if } (1-\alpha)+\frac{\alpha}{M}\leqslant p\leqslant 1.
\end{split}
\right.
\end{equation}
\end{widetext}

For different values of $\alpha$, we plot the accumulated $p$ distribution in Fig.~\ref{fig:dp}. For large $\alpha$, the distribution approaches the jump function, which corresponds to the uniform distribution. 
Notice that the analytical result is based on the global noise $\alpha$, which the numerical simulation given in Fig.~\ref{fig:pt}(b) is based on the gate noise $\epsilon$.  In general, $\alpha$ is a (complicated) function of $\epsilon$ that depends on both $N$ and $D$ (Table I provides some intuition of this function for $N=10,D=24$). Nevertheless, the results of Fig.~\ref{fig:pt}(b) qualitatively agree with the analytical results given by Fig.~\ref{fig:dp}, which also demonstrates that the MPDO simulation delivers reasonable output.

\bibliography{Noise}

\end{document}